\documentclass[12pt,preprint]{aastex}

\def\kmps{\ifmmode {\rm\,km\,s^{-1}}\else 
             ${\rm\,km\,s^{-1}}$\fi}
\newcommand{\kms}{\kmps}
\def\ergcms{\ifmmode {\rm\,ergs\,cm^{-2}\,s^{-1}}\else
             ${\rm\,ergs\,cm^{-2}\,s^{-1}}$\fi}
\newcommand{\fluxunit}{\ergcms}
\def\ergcmsa{\ifmmode {\rm\,ergs\,cm^{-2}\,s^{-1}\,\AA^{-1}}\else
             ${\rm\,ergs\,cm^{-2}\,s^{-1}\,\AA^{-1}}$\fi}
\newcommand{\fluxd}{\ergcmsa}
\def\ergs{\ifmmode {\rm\,ergs\,s^{-1}}\else
             ${\rm\,ergs\,s^{-1}}$\fi}
\newcommand{\lum}{\ergs}
\def\lya{\ifmmode {\mbox{Ly}\alpha}\else
             {Ly$\alpha$}\fi}

\slugcomment{}
\shorttitle{Extended \lya\ Sources at $z\sim 3-5$}
\shortauthors{Saito et al.}

\begin{document}

\title{Systematic Survey of Extended Lyman-$\alpha$ Sources Over $z\sim 3-5$
\footnote{Based on data collected at the Subaru Telescope, 
which is operated by the National Astronomical Observatory of Japan.}}
\author{Tomoki Saito\altaffilmark{2}, Kazuhiro Shimasaku\altaffilmark{2}, 
Sadanori Okamura\altaffilmark{2}, Masami Ouchi\altaffilmark{3,4}, 
Masayuki Akiyama\altaffilmark{5}, 
and Michitoshi Yoshida\altaffilmark{6}}

\altaffiltext{2}{Department of Astronomy, University of Tokyo, 
    7-3-1 Hongo, Bunkyo, Tokyo 113-0033, Japan}
\altaffiltext{3}{Space Telescope Science Institute, 
    3700 San Martin Drive, Baltimore, MD 21210, USA}
\altaffiltext{4}{Hubble Fellow}
\altaffiltext{5}{Subaru Telescope, National Astronomical Observatory of Japan, 
650 North A'ohoku Place, Hilo, HI 96720, USA}
\altaffiltext{6}{Okayama Astrophysical Observatory, 
National Astronomical Observatory of Japan, 
3037-5 Honjo, Kamogata, Okayama 719-0232, Japan}

\begin{abstract}
Spatially extended \lya\ sources which are faint and compact in continuum 
are candidates for extremely young galaxies (age of $\lesssim 10^7\rm yrs$) 
at high redshifts. 
We carried out a systematic survey for extended \lya\ sources, 
using deep intermediate-band imaging data taken with the Suprime-Cam 
on the Subaru Telescope. 
Our survey covers a field of view of $33'\times 25'$ and 
a redshift range of $3.24\lesssim z\lesssim 4.95$, 
down to a \lya\ flux of $\sim 1\times 10^{-17}\fluxunit$. 
We identified 41 extended \lya\ sources. 
The redshift distribution of these sources shows that 
this kind of objects are common in the early universe 
over the surveyed redshift range. 
The identified objects have typical sizes of $\sim 10-15\rm kpc$ 
and luminosities of $\sim 10^{42}\lum$.  
Follow-up spectroscopy made for 7 of the 41 objects 
showed that our sample suffers  from little contamination. 
All 7 objects have large equivalent widths of \lya\ emission line, 
all but one exceeding 240\AA\ in the rest frame. 
The large equivalent widths suggest that their extended \lya\ emissions 
are unlikely to be due to normal starbursts, 
but possibly originated from very young galaxies. 
All 41 objects in our sample have much smaller \lya\ luminosities 
than the two \lya\ Blobs (LABs) found by \citet{s00} in spite of our 
much larger survey volume. 
This suggests that large and luminous extended \lya\ objects like 
the two LABs are very rare and are clustered in overdense regions. 
\end{abstract}



\section{Introduction}
\label{introduction}
It has been said for a long time that galaxies undergoing their very first 
star formation must have strong \lya\ line emission \citep{partridge}. 
Recent large aperture telescopes have conducted many narrowband deep imaging 
surveys, which have successfully identified a large number of \lya\ emitters 
(LAEs) at high redshifts 
(e.g. \citealt{ch98, hu98, fynbo, rhoads00, ouchi03, shimasaku03, 
kodaira, maier, fynbo03, dawson04}). 
These LAEs show a relatively faint continuum and strong \lya\ line emission. 
These surveys provide valuable information about very young galaxies, 
and therefore galaxy formation at high redshifts. 
However, it is not clear whether or not they are really young galaxies, 
i.e., under the initial phase ($\ll 10^7\rm yrs$) of their star formation. 
The LAEs found so far have substantial UV (rest frame) continua, 
though they are relatively faint.
Hence they should already have a certain amount of stellar population. 
This means that most of the known LAEs are evolved galaxies. 
Some LAEs show extremely large equivalent widths, but they can 
be explained with ordinary starbursts by introducing a top-heavy 
IMF or zero-metallicity \citep{mr2002}. 

It is known from many surveys that 
most of the high-$z$ LAEs are relatively compact in \lya\ emission 
(e.g. \citealt{pascarelle, westra}), 
although they are somewhat more extended in \lya\ than in 
continuum (e.g. \citealt{moller98}). 
On the other hand, 
galaxies in early phases ($\lesssim 10^7\rm yrs$) of their formation 
are predicted to have a spatially extended \lya\ line emission originating from 
non-stellar mechanisms such as 
cooling radiation 
\citep{haiman00, fardal, bd03, db06, dijkstra05a, dijkstra05b}, 
resonant scattering of \lya\ photons from central AGNs 
\citep{hr01, weidinger04, weidinger05}, 
or galactic superwinds driven by hidden starbursts 
\citep{tsk01, ohyama, wilman}, 
while they are fairly faint and/or compact in rest-frame UV continuum. 
Accordingly, we will be able to find extremely young galaxies 
by searching for {\em extended} \lya\ sources which are faint and/or 
compact in continuum. 
The first distinctive example of such objects is the two 
Lyman-$\alpha$ Blobs (LABs) found at $z\sim 3.1$ by Steidel et al. 
(2000: hereafter S00). 
The two LABs received strong attention because of the unusual properties. 
Their strong \lya\ emissions (\lya\ luminosities of $\sim 10^{44}\lum$) 
are more clearly extended than found by \citet{moller98}, 
with spatial extents of $\gtrsim 100h^{-1}\rm\,kpc$. 
Despite their extensive \lya\ emission, 
they do not have detectable UV continuum sources sufficient 
to ionize the large amount of ambient \ion{H}{1} gas. 
Although many observational studies ranging from radio to X-ray have 
been made (e.g. \citealt{chapman01, chapman04, ohyama, b04, bower, wilman}), 
their nature is not yet clear. 

It is also totally unknown how commonly extended \lya\ sources 
exist at redshifts other than $z\sim 3.1$. 
There are only a restricted number of extended \lya\ sources known to date 
( e.g. \citealt{keel, francis, dey05, nilsson}). 
Currently the largest sample of extended \lya\ sources is 
the sample of 35 objects by Matsuda et al. (2004: hereafter M04). 
More recently, \citet{nilsson} made a narrowband imaging of a blank 
field (GOODS-S) and identified one object at $z=3.16$ with extended 
\lya\ emission and no detectable continuum counterparts. 
However, their samples are based on narrowband imaging 
which covers only small redshift ranges. 
M04's survey is targeted at a particular overdense region including 
S00's two LABs, and the redshift coverage is fairly small; 
$z= 3.09 ^{+0.04}_{-0.03}$. 
Nilsson et al. has yet smaller redshift coverage; $z=3.157^{+0.017}_{-0.031}$. 
Thus, there is currently no systematic survey in a blank field 
which covers a wide range of redshift. 

Our current work is aimed at searching for extended \lya\ sources 
over a wide redshift range, by performing intermediate-band imaging. 
Since our intermediate-bands have wider bandwidths than usual 
narrowbands used for LAE surveys, 
we can cover a wider range of redshift. 
Although our survey sensitivity is 
restricted to strong emitters having large equivalent widths, 
extended \lya\ sources with faint continuum like S00's LABs 
have sufficiently large equivalent widths to be detected in our survey. 

The rest of this paper is organized as follows. 
We introduce our imaging survey in \S\ref{obs:image}, 
and the photometric selection of extended \lya\ sources in \S\ref{sample}. 
Then we describe the observational details of follow-up spectroscopy for 
the photometrically selected sample in \S\ref{obs:spec}. 
The photometric properties and spectral properties are 
presented in \S\ref{phot} and \S\ref{spec}, 
and a discussion based on their luminosity function is presented 
in \S\ref{LF}. 
Then we discuss currently proposed scenarios for extended \lya\ sources 
in \S\ref{origin} using our imaging and spectroscopic data. 
We summarize our conclusions in \S\ref{conclusion}. 

Throughout this paper, we adopt cosmological parameters 
$\Omega_{\rm M}=0.3$, $\Omega_\Lambda=0.7$, 
and $H_0=100h = 70\kms \rm Mpc^{-1}$. 
All magnitudes are in the AB system \citep{oke, fsi95}.

\section{Photometric sample of extended \lya\ sources}
\label{obs}
\subsection{Intermediate-band imaging survey}
\label{obs:image}
\begin{figure}
\begin{center}
\plotone{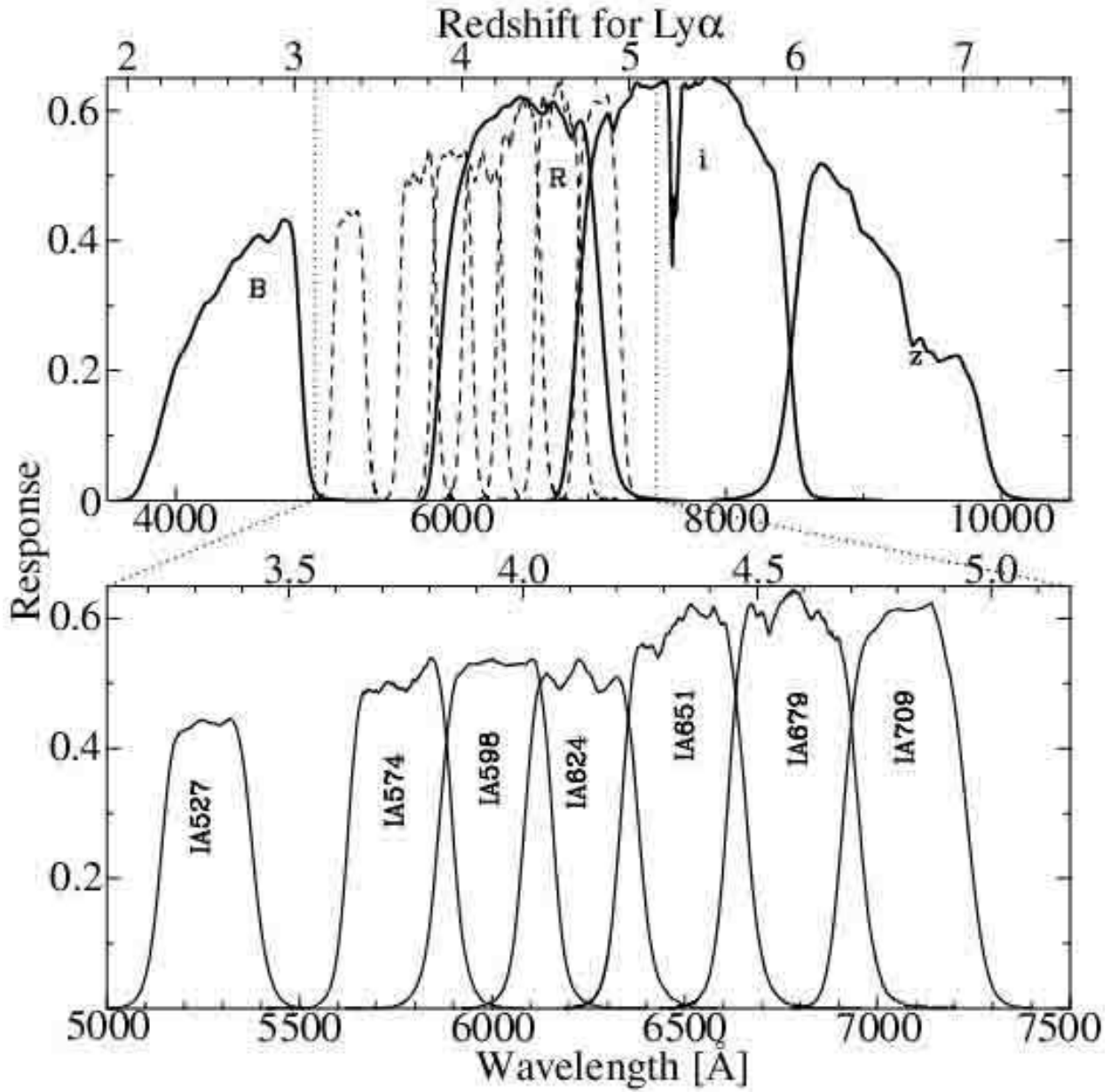}
\caption{
The upper panel shows the responses of the filters used in our survey. 
The dashed curves represents the seven intermediate-band (IA) filters, 
and the solid curves represent the four broad-band filters. 
The lower panel is a zoom up for the IA filters. 
The abscissa represents the wavelength (marked on the bottom) 
and the redshift for the \lya\ line (marked on the top). 
}
\label{fig:filters}
\end{center}
\end{figure}

We carried out a wide-field, deep imaging survey with the 
{\it Subaru Prime-focus Camera} (Suprime-Cam: \citealt{miyazaki}) 
mounted on the 8.2m Subaru Telescope \citep{iye}, 
using seven intermediate-band filters 
(IA filter system: \citealt{hayashino}). 
The response curves of our filter set are shown in Fig.\ref{fig:filters}. 
Each filter has a bandwidth of $\sim 240-340\rm\AA$ 
giving $R = \lambda/\Delta\lambda \sim 23$. 
We can detect \lya\ emission from almost the whole redshift 
range between $z=3.24$ and $z=4.95$ using these seven filters. 
The Suprime-Cam has a wide field of view, $34'\times 27'$ 
(central $33'\times 25'$ was used in this survey), 
so that our survey covers a large volume of 
$\sim 1\times 10^6h^{-3}\mbox{Mpc}^3$ in total. 
The observations were made on October 30 and 31 and November 6 in 2002 
as an open-use program of the Subaru Telescope (S02B-163: Kodaira et al.). 
We pointed the telescope toward a blank-field called 
the Subaru/XMM-Newton Deep Field South (SXDF-S), which is 
located at $\alpha = 02^h18^m00^s$, $\delta = -05^\circ 25' 00''$ (J2000). 
Relevant data for the intermediate-bands 
are summarized in Table \ref{tab:img}. 
The $3\sigma$ limiting magnitude ranges from 26.5 to 26.8 mag ($2''\phi$ 
circular aperture). 

The SXDF-S has deep Suprime-Cam images in the $B$, $R$, $i'$, 
and $z'$ bands. These broadband images were obtained in 2002 and 2003 
in a large project of the Subaru Telescope named Subaru /XMM-Newton 
Deep Survey (SXDS: \citealt{sekiguchi05}). 
The exposure times and limiting magnitudes for the broadband images are 
also summarized in Table. \ref{tab:img}. 
The $3\sigma$ limiting magnitudes in a $2''\phi$ circular aperture are 
28.2 ($B$), 27.4 ($R$), 27.0 ($i'$), and 25.8 ($z'$) mag, respectively. 
The response curves of these broadbands are also shown in 
Fig.\ref{fig:filters}. 
Both the intermediate-band and broadband data were reduced with 
our own software tools based on NOAO IRAF software package.  
The photometric zero-points were calculated with $\ge 100$ bright stars 
within the FoV of the Suprime-Cam, and corrected by amount of $\simeq 0.05$ 
mag to match their colors to those of galactic star spectra provided 
by \citet{gs83}. 
The seeing sizes ranged from $0''81$ (broadbands) to $1''.05$ 
(IA679 \& IA709) in the raw images. 
All the images were smoothed with a Gaussian kernel 
to match their PSF sizes to the worst-seeing image ($1''.05$). 
After matching the PSF sizes, we performed source-extraction 
and photometry using SExtractor ver. 2.1.6 \citep{bertin}. 
We used the broadband data to measure continuum levels in selecting 
LAEs. 
Since the Suprime-Cam has a very wide FoV, our images have a few 
extremely bright stars and their ghost images. 
We carefully defined the regions affected by these stars and ghosts 
as well as a number of saturated stars, 
and did not use the sources extracted in these regions for further 
analysis. We also rejected the regions near the edge of the images. 
The regions rejected here make up $\sim 4.2\%$ of the whole area.

\subsection{Photometric selection of extended \lya\ sources}
\label{sample}
We have constructed a photometric sample of candidate extended \lya\ 
sources from the intermediate (IA)-band and broadband imaging data. 
We measured the magnitude basically with a $2''\phi$ circular 
aperture unless otherwise noted. 
We selected candidates from objects brighter than $5\sigma$ in 
the IA-detection catalogues created with SExtractor. 
The outline of the selection of extended \lya\ sources is as follows. 
We first selected LAE candidates satisfying the following two criteria 
simultaneously: 
(1) the \lya\ emission line has a sufficiently large equivalent width, 
and (2) non-detection or a large dropout in the $B$ band. 
Then we further imposed the following two criteria on the sources 
selected with (1) and (2): 
(3) the spatial extent is significantly larger than point sources 
in the IA band, 
and (4) compact or faint in the continuum band red-ward of the \lya\ line. 

First, we selected objects of which 
the \lya\ line has a sufficiently large equivalent width to be 
detected as IA-band excess relative to the broadband continuum strength. . 
The continuum levels to estimate equivalent widths (``Cont'') 
were measured with the broadband data red-ward of 
(and the closest to) the \lya\ line, 
i.e., $R$-band for IA527 and IA574, $i'$-band for IA598, IA624, IA651, 
and IA679, and $z'$-band for IA709.  We adopt the criterion of 
$\mbox{Cont}-\mbox{IA} > 0.75\mbox{mag}$, which corresponds to 
an observed-frame equivalent width $EW_{obs}\gtrsim 240{\rm \AA}-340\rm\AA$ 
($\gtrsim 55\rm\AA$ in the rest-frame). 
Such large equivalent widths are rarely observed in foreground 
[\ion{O}{2}] emitters, and the selected objects are likely to be 
\lya\ emitters at high redshifts. 
In order to measure the total \lya\ line flux, 
we used the IA magnitude measured with an automatic aperture 
(MAG\_AUTO of SExtractor) in calculating the IA excess. 
On the other hand, we used the continuum magnitude measured with 
a $2''\phi$ circular aperture to minimize the effect of sky noise. 
This mixed use of different apertures enables us to effectively 
collect the \lya\ line flux, and to measure the continuum level 
with a deeper limiting magnitude. 
Although this may increase contaminants due to its systematic 
overestimate of the IA excess, 
we can exclude contaminants with the fourth criterion, 
``compact or faint in continuum'', as described below. 

Second, we selected from the IA-excess objects only those that 
were ``drop out'' or very faint ($< 2\sigma$) in the $B$ band. 
This means that we selected objects with a large continuum 
gap between the $B$ band and the continuum red-ward of the \lya\ line. 
For the two bands corresponding to the lowest redshifts (IA527 and IA574), 
the expected continuum gap is not very large and hence objects 
with Lyman discontinuity can be detected even in the $B$ band. 
Therefore, we used another criterion for these two bands: 
$B-R\ge 0.6$ for IA527, and $B-R\ge 1.4$ for IA574. 
Together with the first criterion, large IA-excess, 
we can expect that the selected sample of LAEs contains few 
low-$z$ interlopers. 
These selection criteria are summarized in Table.\ref{tab:select}. 

\begin{figure}
\begin{center}
\plotone{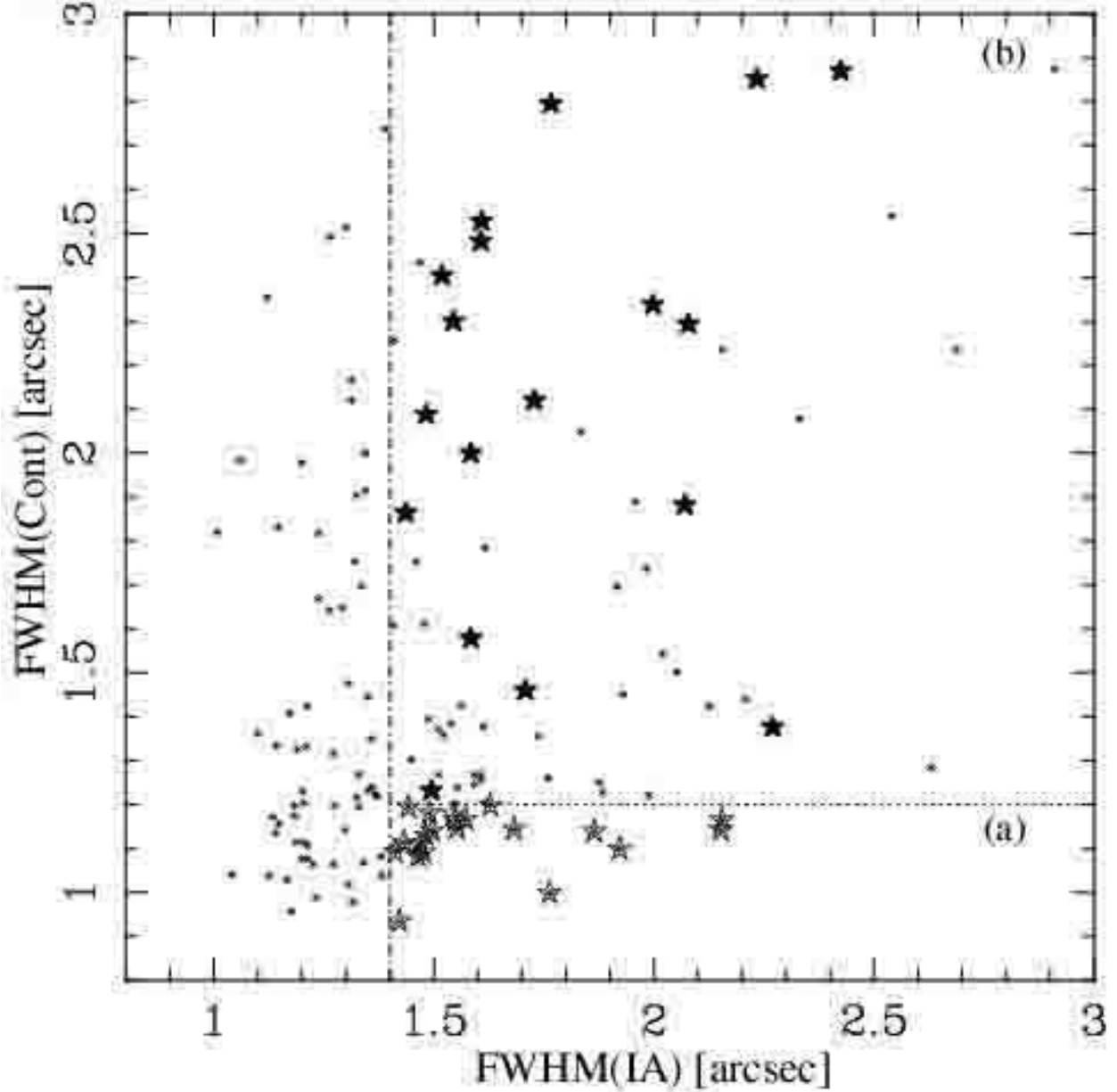}
\caption{Size distribution of the photometrically selected extended \lya\ 
source candidates in the IA and continuum bands. 
The abscissa shows FWHM in the detected IA band, and the ordinate shows 
FWHM in the continuum band red-ward of the \lya\ line. 
The asterisks indicate all objects satisfying the criteria of \lya\ 
excess and $B$-band non-detection/dropout. 
The star marks represent our sample of extended \lya\ sources. 
The open stars in region (a)  are continuum-compact objects. 
The filled stars in region (b) are continuum-faint objects, which 
are too faint in continuum to have the sizes measured with a 
reasonable accuracy, i.e., the abscissa values for these objects 
are not reliable (see the text). 
The threshold for the spatial extent in \lya\ is shown with the 
vertical dot-dashed line, and that in continuum 
is shown with the horizontal dotted line. 
}
\label{fig:fwhm-IBcont}
\end{center}
\end{figure}

\begin{figure}
\begin{center}
\plotone{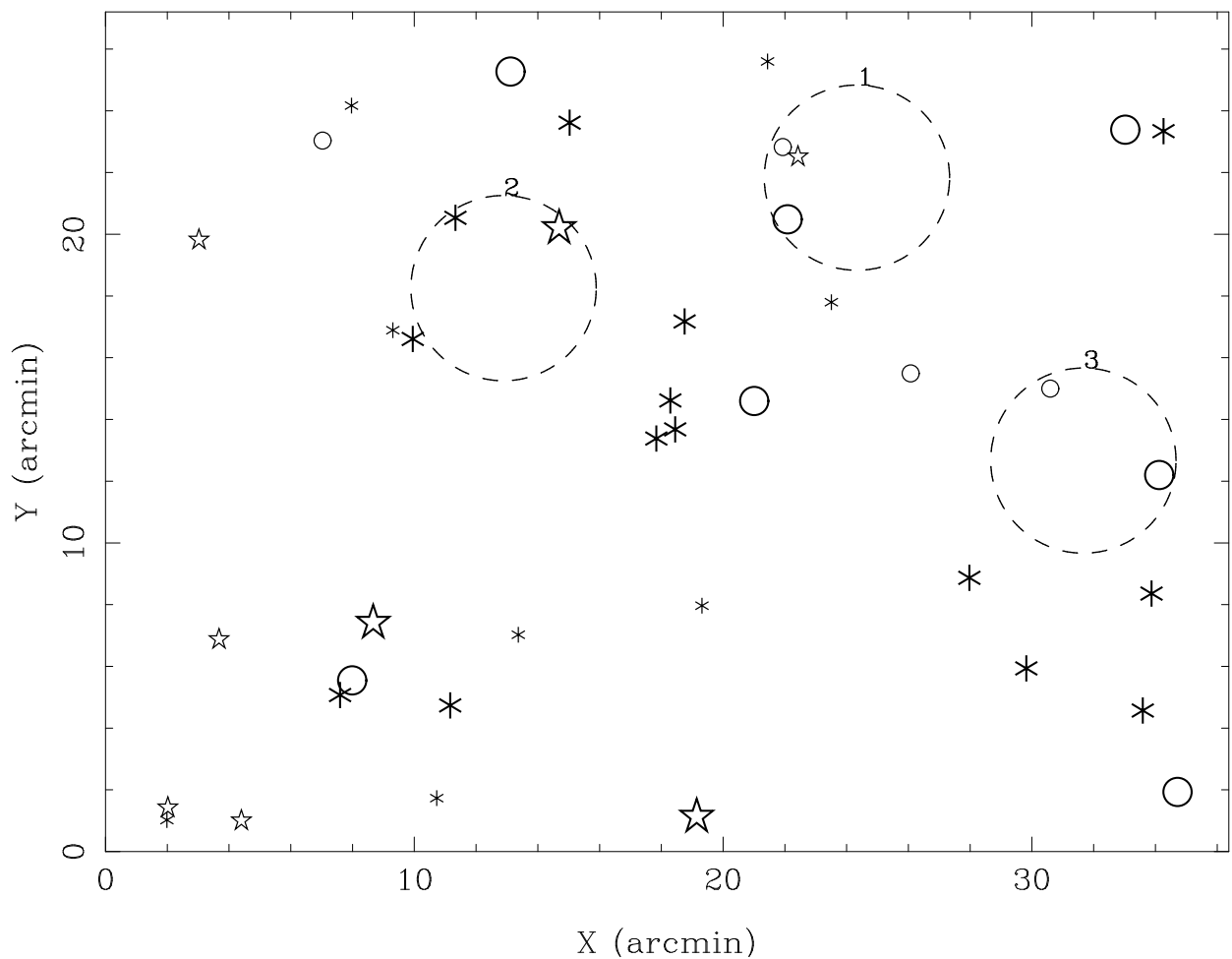}
\caption{Sky distribution of extended \lya\ source candidates 
in our sample. 
The dashed circles indicate the three MOS FoVs of follow-up spectroscopy. 
Spectra were obtained for 7 of the 41 candidates. 
The open stars and open circles indicate objects with 
a large \lya\ excess: $\mbox{Cont}-\mbox{IA}>2.5$ and 
$\mbox{Cont}-\mbox{IA}>1.5$, respectively, where 
the Cont is the continuum magnitude measured in the broadband 
red-ward of the \lya\ line. 
Objects marked with large symbols are brighter than 
25.5 mag in the IA band. 
}
\label{fig:skydist}
\end{center}
\end{figure}

Then, we selected spatially extended \lya\ sources from the LAE sample 
selected above. The spatial extent of \lya\ line component was measured 
in the detected IA-band, and that of continuum component was measured 
in the broadband where we measured the continuum level for estimating the 
\lya\ excess. 
The criterion we adopted for ``extended'' is FWHM(IA)$>1''4$, 
which means that the spatial extent of an object is larger than 
the PSF size ($1''.05$) by more than $\simeq 2$ pixels ($0''.4$). 
As shown in Fig.\ref{fig:fwhm-IBcont}, we find that there are two kinds 
of objects among these ``extended'' \lya\ sources. 
One is continuum-compact objects, that is, 
``objects which can be regarded as point-sources in continuum'' 
(those in (a) in Fig.\ref{fig:fwhm-IBcont}), 
and the other is continuum-faint objects, that is, 
``objects which are too faint in continuum to reliably measure the 
FWHM in the continuum images''
(those in (b) in Fig.\ref{fig:fwhm-IBcont}). 
We regarded an object as ``continuum-compact'' if 
it is brighter than the $3\sigma$ sky noise in the continuum band but 
its size is smaller than $1''.2$ (FWHM), 
and as  ``continuum-faint'' if it is 
fainter than $3\sigma$ in the continuum band. 
Both groups are included in our sample because they all have 
extended strong \lya\ emission. 
We selected 43 objects following these criteria; 
23 continuum-compact objects and 20 continuum-faint objects. 
There are two double-counted objects, however, whose redshifts 
are covered by two IA bands; IA598-IA624, and IA651-IA679. 
After eliminating these two double counts, we finally obtained 
a sample of 41 extended \lya\ sources. 
This sample includes 22 continuum-compact objects and 19 continuum-faint 
objects. 
The objects selected here are analogous to the two LABs of S00, 
but objects with smaller spatial extents are included 
because we adopted a smaller limiting size than the size of the LABs. 
The sky distribution of the 41 objects is shown in 
Fig.\ref{fig:skydist}. 

The magnitudes and FWHM sizes of our objects 
selected above are listed in Table.\ref{tab:phot}. 
As already mentioned, 
the magnitudes were measured basically with a $2''\phi$ circular 
aperture, but automatic aperture magnitudes (MAG\_AUTO) 
were used for the IA band where \lya\ emission was detected. 
The magnitudes in the IA band where \lya\ line was detected and 
the nearest broadband red-ward of the \lya\ line are written 
in bold characters in Table.\ref{tab:phot}. 
Although we did not put any constraints on the continuum gap between 
the $B$-band and the red-ward broadband of \lya\ for the bands redder 
than IA598, we found that 
all objects in our sample have fairly red $B-\mbox{Cont}$ colors.

\section{Follow-up spectroscopy}
\label{obs:spec}
We carried out follow-up spectroscopy for seven objects 
among the 41 objects selected above. 
The observations were made on 2003 November 27, using {\it Faint Object 
Camera And Spectrograph} (FOCAS: \citealt{focas}) mounted on the 
Cassegrain focus of the Subaru Telescope. 
We used the multi-object spectroscopy (MOS) mode with three MOS masks, 
and obtained spectra of the seven candidate objects. 
The sky distribution of these MOS fields is also shown in 
Fig.\ref{fig:skydist}. 

The 300B grism, the SY47 order-cut filter, and $0''.8$-wide slits were 
used, 
giving a wavelength coverage of $\lambda\sim 4700-9100\rm\AA$. 
This is sufficient to detect [\ion{O}{3}] $\lambda\lambda$4959, 5007 
emissions in case the object is a foreground [\ion{O}{2}] $\lambda$3727 
emitter. 
The spectral resolution of this setting is $R\simeq 500$. 
This nominal resolution is equivalent to a wavelength resolution of 
$\simeq 12\rm\AA$ at $\lambda=6000\rm\AA$, 
and a velocity resolution of $\simeq 600\kms$. 
The resolutions of spectra we finally obtained depend on the number 
of combined frames and the distance from the sky emission lines used 
for the references in the wavelength calibration. 
We measured the velocity width of the sky emission lines nearest 
to the \lya\ line, and estimated the 
velocity resolutions to be $\sim 670 - 800\kms$. 
The exposure time for each mask was 1.5-2.5 hrs, giving 
an rms noise level of $\sim 10^{-19} \fluxd$ (25-26 mag). 

\begin{figure}
\begin{center}
\plotone{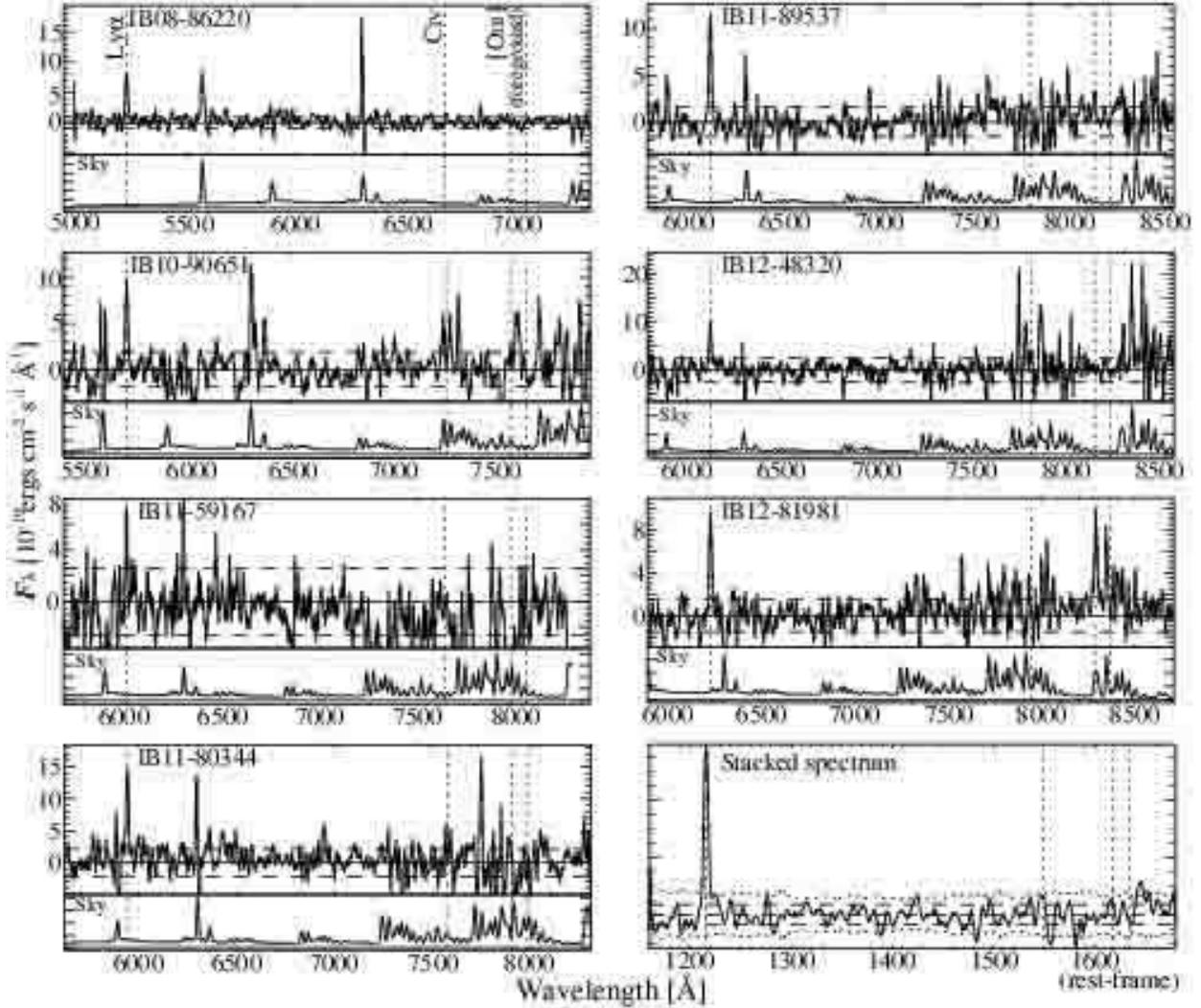}
\caption{
One-dimensional spectra of the seven extended \lya\ sources, 
integrated over the full extents along the slit direction. 
The wavelength ranges plotted here are $1150\le \lambda/(1+z) \le 1700\rm\AA$. 
The wavelengths of \lya, \ion{C}{4}, and foreground [\ion{O}{3}] 
are marked with the vertical dotted lines. 
The horizontal dashed lines are $\pm 1\sigma$ noise levels 
calculated from the whole spectral range plotted here. 
The sky spectrum is plotted at the bottom of each panel. 
All the wavelengths are in the observed-frame, except for the stacked 
spectrum shown in the bottom-right panel. 
For the stacked spectrum, the dotted curves show $\pm 1\sigma$ noise levels 
calculated from the photon counting errors. 
}
\label{fig:spectra}
\end{center}
\end{figure}

The one-dimensional spectra of the seven objects are shown in 
Fig.\ref{fig:spectra}, 
and their spectral properties are summarized in Table \ref{tab:spec}. 
The wavelength range plotted in Fig.\ref{fig:spectra} 
covers both the [\ion{O}{2}] line and [\ion{O}{3}] doublet lines in 
case the object is an [\ion{O}{2}] emitter at redshift 
$0.38\lesssim z \lesssim 0.94$. All of the spectra show 
only one strong emission line within the range of the detected IA bands 
except for spurious ones due to imperfect sky subtraction. 
None of them shows a detectable 
continuum emission down to $\sim 10^{-19}\fluxd$. 
We therefore conclude that all of them are single line objects, 
and that they all have large equivalent widths, as anticipated 
from the photometry. 
Similar results can be seen in the stacked spectrum shown in the 
bottom-right panel of Fig.\ref{fig:spectra}. 
Neither significant continuum emissions nor emission lines other than the 
candidate \lya\ can be seen even in the stacked spectrum. 

\begin{figure}
\begin{center}
\plotone{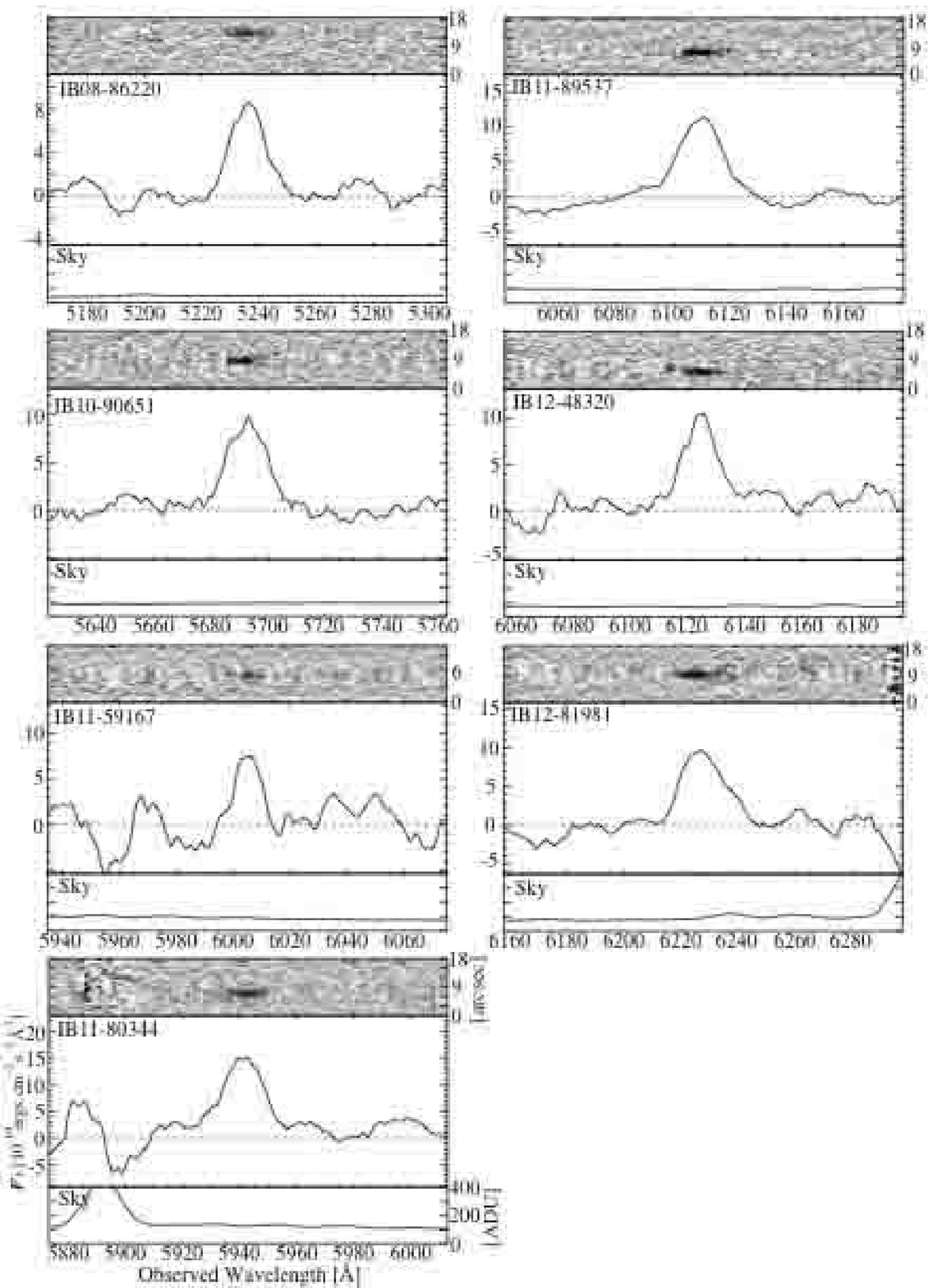}
\caption{Two-dimensional and one-dimensional spectra around 
the \lya\ emission line of the seven extended \lya\ sources. 
The horizontal dotted lines represent the zero-level. 
The wavelength ranges plotted are $\pm 70\rm \AA$ of the central 
wavelengths. 
The corresponding sky spectrum is plotted at the bottom of each panel. 
}
\label{fig:spec-line}
\end{center}
\end{figure}

It is clear that 
three objects in our sample, IB08-86220, IB11-89537, and IB12-48320, 
do not have [\ion{O}{3}] emission at the expected wavelength if the 
candidate \lya\ line is in reality a foreground [\ion{O}{2}]. 
If they are foreground [\ion{O}{2}] emitters, there should be 
[\ion{O}{3}] line emission detectable with our current spectral data. 
We therefore conclude that they are LAEs at high redshifts. 
For the rest of the sample, the spectra around the expected 
wavelength of the foreground [\ion{O}{3}] lines are overlapped with 
sky emission lines. 
The emission lines detected in there are almost completely consistent with 
residuals of the known sky spectrum due to imperfect sky subtraction. 
Fig.\ref{fig:spec-line} shows the spectra around the candidate \lya\ line. 
One object in our sample, IB12-81981, 
shows a marginal signature of an asymmetric line profile with a 
relatively sharp blue cut-off and a broader red wing, 
which are signatures of \lya\ lines at high redshifts. 
The line profiles of the another two objects, IB11-89537 and IB12-48320, 
also show a marginal signature of asymmetry. 
It should be noted here that all seven objects 
have large equivalent widths of the candidate \lya\ line. 
As described in \S\ref{spec}, 
even the object with the smallest equivalent 
width, IB08-86220, has $EW_{rest}\sim 100\rm\AA$ if the line is \lya . 
If this is a foreground [\ion{O}{2}] emitter at $z=0.405$, 
the $EW_{rest}$ is as large as $300\rm\AA$, which is hardly explained by 
an ordinary starburst galaxy. 

By taking three pieces of observational information, 
detection of a single line, asymmetric line profile, 
and large equivalent width, we conclude that at least four 
objects, and probably all seven objects, are high-$z$ \lya\ emitters. 

\section{Results and discussion}
\subsection{Photometric properties}
\label{phot}
\begin{figure}
\begin{center}
\plotone{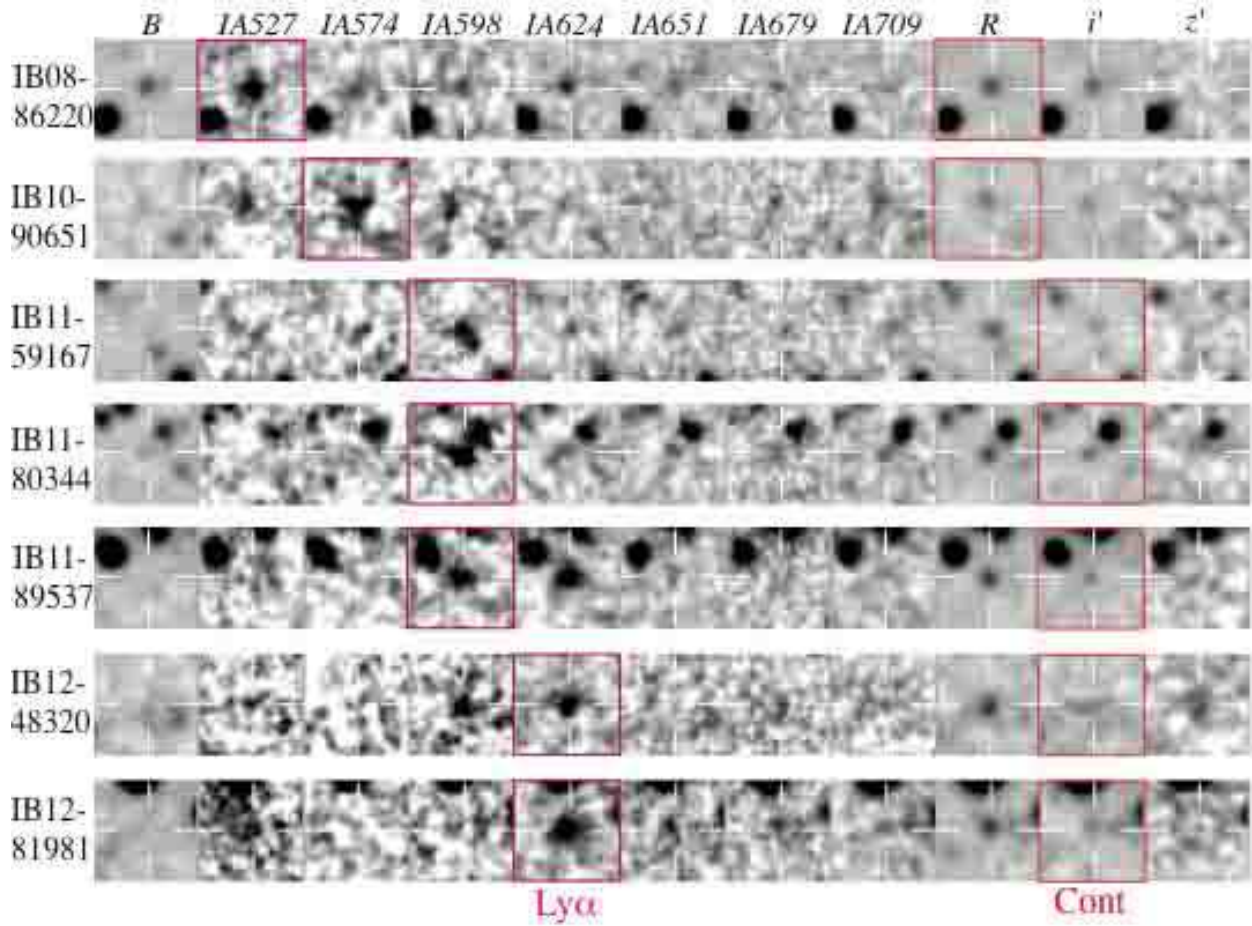}
\caption{Images of the seven extended \lya\ sources for which 
follow-up spectroscopy was made. 
The detected IA band images are outlined with magenta, 
and the broadband images used for measuring the red-ward continuum 
level are outlined with red color. 
Each image shows a region of $6''\times 6''$. 
}
\label{fig:portraits}
\end{center}
\end{figure}

As seen in Table.\ref{tab:phot}, 
all 41 extended \lya\ sources of our sample are relatively 
faint in red-ward continuum. 
The spatial extents in continuum for 22 objects are clearly compact. 
The remaining 19 objects could be extended in continuum but 
their sizes are not measured reliably because they are too faint. 
The images of the seven objects observed spectroscopically are 
shown in Fig.\ref{fig:portraits}. 

\begin{figure}
\begin{center}
\plotone{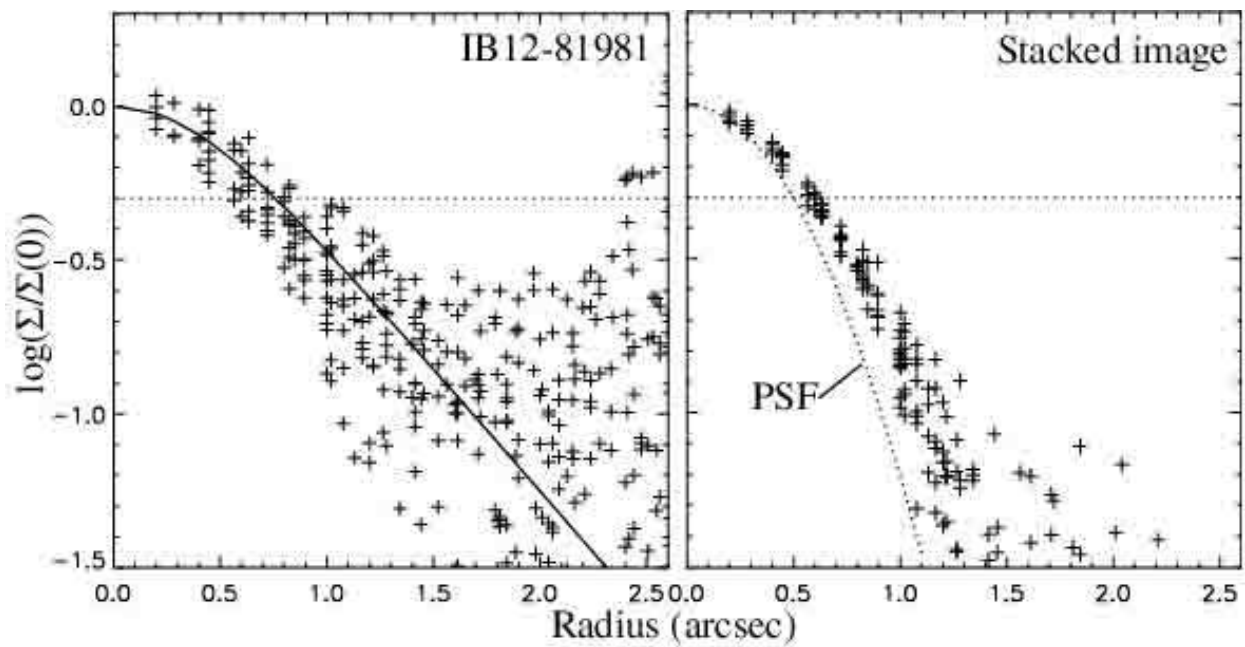}
\caption{({\it Left}) Radial surface brightness profile of IB12-81981, 
a typical object in our sample. 
Normalized surface brightness is plotted against the distance 
from the center of the object. The crosses show the surface brightness 
of each pixel, and the solid curve is the radial profile 
of a PSF-smoothed exponential disk with an appropriate half-light radius (HLR).
({\it Right}) Radial surface brightness profile of the stacked image of 
the 41 objects. The dotted curve shows the Gaussian PSF with FWHM of $1''.05$. 
For both panels, the horizontal dotted line shows a half level of the 
central surface brightness. 
}
\label{fig:rprofile}
\end{center}
\end{figure}

\begin{figure}
\begin{center}
\plotone{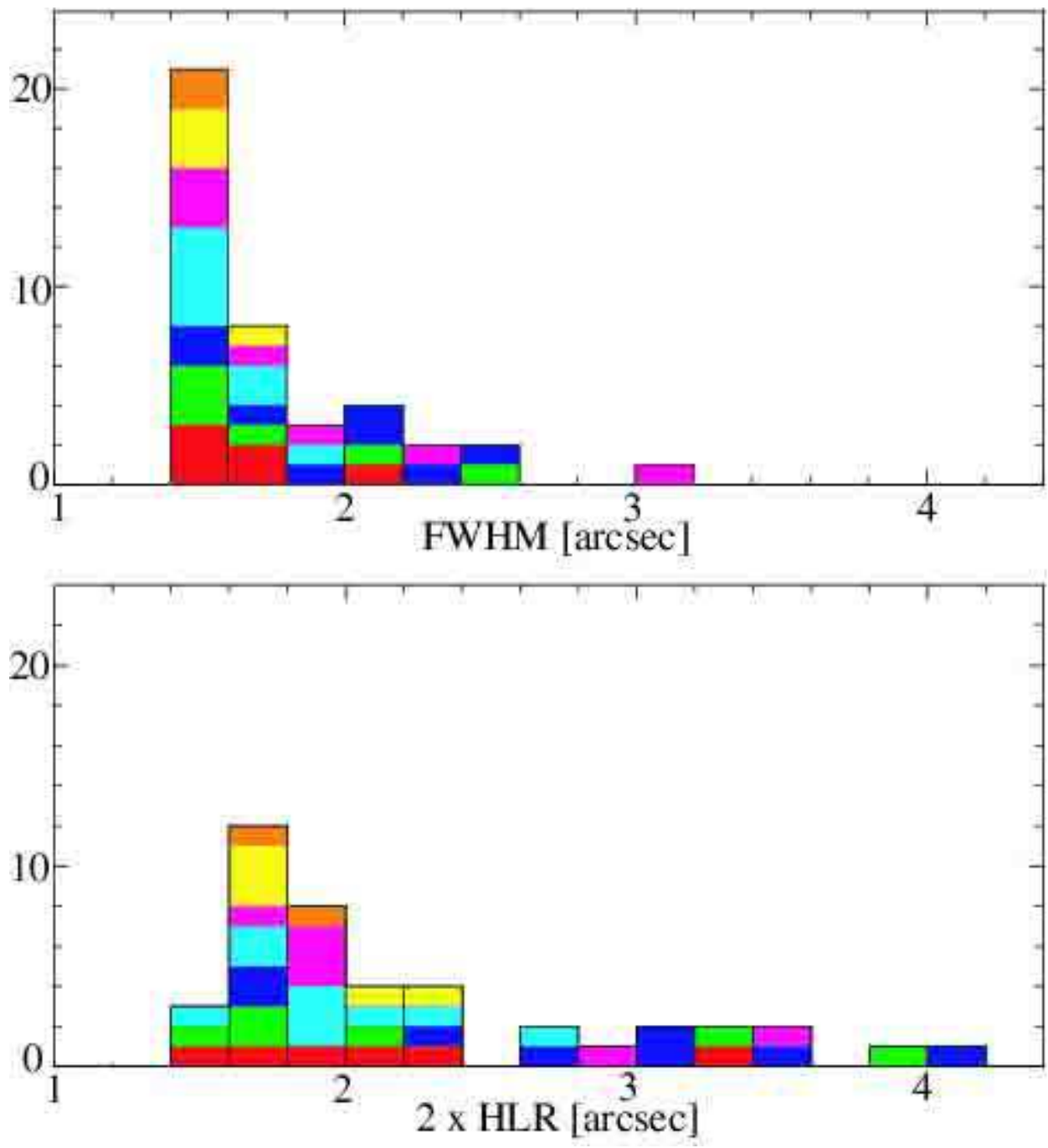}
\caption{Size distribution of extended \lya\ sources.
The top panel is for the FWHM sizes measured on the images. 
The bottom panel is for the intrinsic size defined as twice the 
half-light-radius (HLR), 
calculated in our imaging simulation (see text for details). 
The colors show the detected IA bands: red for IA527, green for IA574, 
blue for IA598, cyan for IA624, magenta for IA651, yellow for IA679, 
and orange for IA709. 
}
\label{fig:sizedist}
\end{center}
\end{figure}

In order to estimate the true spatial extents, 
we made artificial images of PSF-smoothed exponential disks with 
various sizes and measured their FWHMs. 
We then established an empirical relation between the intrinsic 
half-light radii (HLR) of the exponential disk and the measured FWHMs of 
their PSF-smoothed artificial images. 
The HLRs of our objects were estimated using this FWHM-HLR relation. 
Although we do not know the precise radial profiles and shapes of our objects, 
exponential disks look a reasonable model. 
The left panel of Fig.\ref{fig:rprofile} shows the radial profile of a 
typical object in our sample, IB12-81981, together with the PSF-smoothed 
profile of a best-fit exponential disk. 
The distributions of FWHMs and calculated sizes ($2\times\mbox{HLR}$) are 
shown in Fig.\ref{fig:sizedist}. 
Their typical sizes are $\sim 1''.5$ ($\sim 9.5 - 11\,\rm kpc$ for 
surveyed redshifts) in FWHM, 
and their intrinsic HLRs calculated above are typically 
$\sim 0''.9$ ($\sim 5.7-6.8\,\rm kpc$). 
The physical size, defined as twice the HLR, is typically 
$\sim 1''.8$ ($\sim 11 - 14\,\rm kpc$). 
Strictly speaking, these values are spatial extents of both the 
\lya\ line and continuum emission, because we did not subtract 
the continuum emission in measuring the FWHM. 
However, the FWHMs we obtained are almost solely determined from the 
extents of \lya\ line emission because of their large equivalent widths. 

The colors in Fig.\ref{fig:sizedist} represent detected IA bands, 
i.e., redshifts. 
They seem to be almost uniformly distributed over the surveyed 
redshift range $3.24\lesssim z \lesssim 4.95$, 
showing that this kind of objects are common in the early universe. 
We did not detect any significant redshift dependence of their 
spatial extent. 
Their spatial extents are much smaller than those of S00's LABs which 
exceed $\sim 100\,\rm kpc$. 
The extent of the largest objects in our sample is 
$3''.2$ ($\sim 21\,\rm kpc$) in FWHM and $2''.8$ ($\sim 19\,\rm kpc$) in HLR. 
Because this object (IB13-104299) is, however, affected by a neighbouring 
bright object, we cannot obtain a reliable value of its physical size. 
The second largest object (IB11-3727) has an FWHM of $2''.5$ 
($\sim 17\,\rm kpc$) and an HLR of $2''.0$ ($\sim 14\,\rm kpc$). 
Therefore, even the largest objects in our sample are extended 
less than $\sim 30\,\rm kpc$. 

It is possible that the noise of the images causes systematic 
overestimates of the spatial extents of our objects. 
We stacked the images of the 41 objects, and investigated the radial profile. 
The right panel of Fig.\ref{fig:rprofile} shows the surface brightness 
on the stacked IA image as a function of the radius. 
The radial profile shows that the stacked image is actually more extended 
than the PSF. We also stacked the red-ward continuum images of our objects, 
but found no clear signature that it is more extended than the PSF. 
We then examined how the noise of the observed images 
affects the apparent sizes of our objects by performing 
a Monte-Carlo simulation. Here we used the magnitude data of $B$ band, 
IA-band (detection), and the broadband red-ward of \lya\ line 
(continuum) shown in Table.\ref{tab:phot}. 
The simulation procedure was as follows: 
(1) make 1025 artificial point-sources which have the same magnitude 
distribution as our sample, 
(2) add the Poisson noise, (3) put them randomly onto the original images, 
and (4) re-extract them with the same manner as for the observed data. 
These processes were repeated 100 times, and 
up to $\sim 90\%$ of the artificial objects are successfully re-extracted. 
We counted the re-extracted objects which satisfy our selection criteria 
presented in \S\ref{sample}. 
Our simulation showed that $\sim 97\%$ of the re-extracted artificial 
objects (point sources) did not satisfy the selection criteria of 
our sample. 
This result shows that most of the objects in our sample are 
actually extended, and are not likely to be compact sources which are 
apparently extended due to noise.

Although most of our objects are really ``extended'' sources, 
we must note that our survey has a relatively low sensitivity 
to the \lya\ line due to the wide bandpasses of the IA filters. 
This implies that we cannot detect low surface 
brightness components of \lya\ emitting gas 
and that the sizes we obtained tend to be smaller than real. 
Moreover, because the distances of our objects are larger than that of 
S00's LABs, the size distribution also suffers from cosmological 
dimming. 
For instance, when we detect LAB1/LAB2 with our surface-brightness 
thresholds, their sizes come down to $\sim 10''$ or less in the IA527 image, 
and $\sim 5''$ or less in the IA709. 
To estimate these effects in detail, we performed an imaging simulation, 
using the narrowband images of 35 LABs at $z\simeq 3.1$ provided by M04. 
The procedure of this imaging simulation is as follows. 

We made simulated images of the LABs as if they were 
located at the redshifts covered by our survey and 
were observed through the same IA filters. 
First, we took the cosmological dimming effect into account. 
Next, we lowered the surface brightness of the simulated 
images by the amount corresponding to the ratio of 
the bandwidth of our IA filters to that of M04's filter. 
Our IA filters have bandwidths of 240-340\AA, while M04's has 77\AA.
The surface brightness of the \lya\ line emission was therefore 
lowered by a factor of $4 - 17$ on the simulated images. 
Then we added noise to make the S/N ratios of the final simulated images 
equal to those of our observed IA images. 
We performed source extraction and photometry on the images with 
the same manner as for the observed IA images. 

\begin{figure}
\begin{center}
\plotone{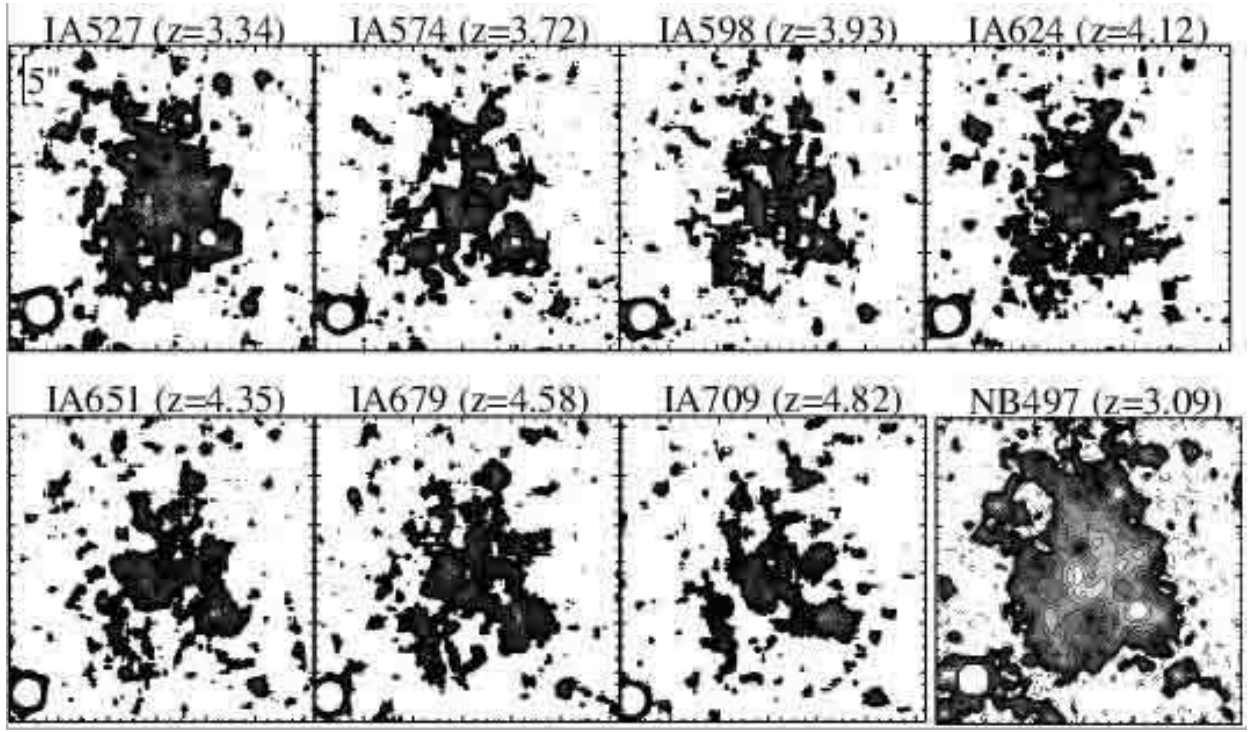}
\caption{Simulated images of the LAB1 located within our redshift range 
and observed with our IA filters.
The bottom-right panel shows the original NB497 narrowband image of 
the LAB1 (Matsuda: private communication). 
The simulated redshifts and corresponding IA bands are marked 
on the top of each panel. For the simulated IA images, the lowest 
contour represents the $2\sigma$ noise level, 
and the contours are marked with an interval of $1\sigma$. 
For the NB497 image, the lowest contour level is $2\sigma$, 
and each contour level is $\sqrt{2}$ times the previous level. 
}
\label{fig:LABsim}
\end{center}
\end{figure}

\begin{figure}
\begin{center}
\plotone{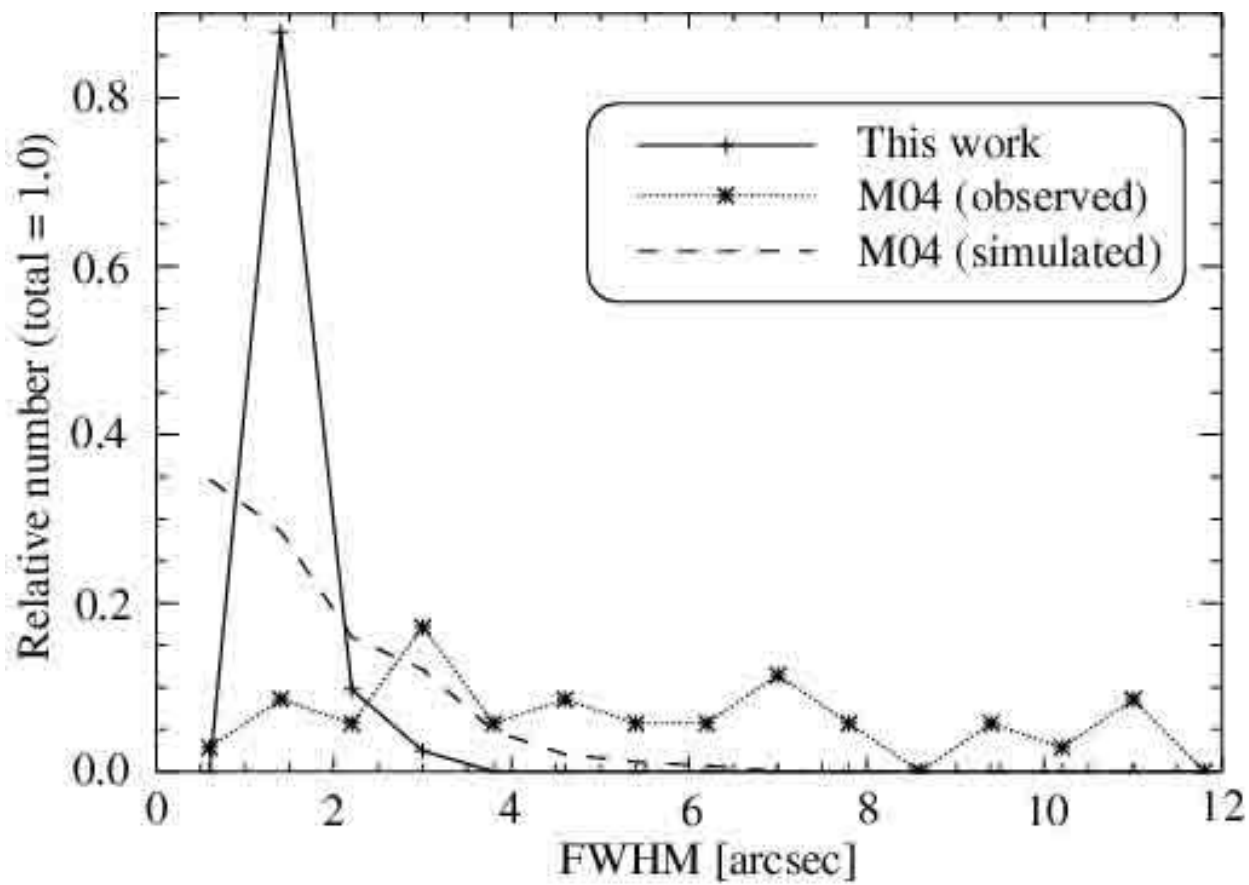}
\caption{
Size distributions of our sample and simulated and original M04 samples. 
The intermediate-band imaging simulation shown in Fig.\ref{fig:LABsim} 
was made for all the objects in M04's sample, and the sizes were 
measured with the same manner as for our data (see the text). 
The size distribution obtained in the imaging simulation is shown with 
the dashed line. This is the averaged distribution of the seven 
simulated images corresponding to the seven IA bands. 
The observed size distribution of our sample is shown with the 
solid line with crosses, 
and that of M04's sample is shown with the dotted line with asterisks. 
}
\label{fig:simsize}
\end{center}
\end{figure}

The simulated images for the LAB1 are shown in Fig.\ref{fig:LABsim}, 
and the simulated size distribution of M04's sample is shown in 
Fig.\ref{fig:simsize}. 
Even the most prominent object, S00's LAB1, gives a much smaller apparent 
size. For example, if it is located at $z=4.82$ and observed through 
the IA709 filter, the measured FWHM will be $\sim 5''$ or so 
($\sim 30\,\rm kpc$). Similarly, if it is located at $z=3.34$ and 
observed through the IA527 filter, the FWHM will be $\sim 10''$ 
($\sim 70\,\rm kpc$). 
Because its diffuse component becomes fainter than the sky 
fluctuation level, the apparent size of the extended emission itself 
becomes smaller, and the extended plateau emission is divided into 
two or more objects. 

Similarly, we found that M04's objects are not much larger than the 
objects in our sample. 
Figs. 10 and 11 compare the sizes of our objects with those of M04's. 
While the M04 objects have much larger observed sizes 
(up to $\simeq 12''$) than our objects ($\lesssim 3''$), 
their sizes will decrease to $\lesssim 6''$ if they are observed 
through the IA filters under the same conditions as for our imaging. 
It should, however, be interesting to note that the average size of the 
``simulated'' images of the M04 objects is slightly larger than 
that of our objects (see Fig. 10). 
The distribution of the simulated sizes of the M04 objects extends 
up to $6''$, while no object in our sample has a size larger than $4''$. 
This trend may suggest environmental dependence of extended \lya\ 
emitting sources. 
The M04 objects reside in the highly overdense region in which S00's 
two large LABs are found. 
Overdense regions may include larger extended \lya\ sources 
than normal fields. 
In fact, large objects like the two LABs have not been found in 
narrowband surveys. 
If such extended objects exist in our field, the surface brightness 
should be lower than the two LABs of S00.

\subsection{Properties of the spectroscopic sample}
\label{spec}
We estimated the \lya\ line fluxes, $F(\lya)$, of the objects in the 
spectroscopic sample from a combination of spectral and photometric data. 
The line fluxes directly obtained from the spectral data 
are smaller than the total $F(\lya)$ values, 
since we used $0''.8$-wide slitlets. 
By making an artificial exponential disk image with the typical 
size of our objects, we calculated the typical fraction of the 
flux collected with a $0''.8$-wide slit to be $\sim 40\%$ of 
the total flux. 

In order to evaluate this fraction independently, 
we estimated the fraction using the broadband photometric data. 
First we assumed that the intrinsic UV continuum is flat within the 
detected IA band, and that this continuum level is equal to 
that of the red-ward continuum estimated from the broadband photometric data. 
Then we made a model flat-spectrum covering the IA band, 
and calculated the absorption of the IGM according to \citet{madau95}, 
using the redshift value obtained from the spectroscopy. 
The pure \lya\ line flux was estimated simply by subtracting the 
absorption-corrected continuum flux from the total flux measured with 
the IA photometry. We compared this photometrically estimated line flux (a) and 
the line flux measured simply by integrating the observed spectrum (b). 
The ratios (b)/(a) calculated for all seven objects in the spectroscopic 
sample ranged from 0.35 to 0.14, which is sufficiently 
smaller than the value for the artificial exponential disk, $\sim 0.4$. 
Because we do not know accurate radial profiles and spatial extents, 
we adopted a conservative conversion factor (b)/(a)$= 0.4$ 
for all the objects in the spectroscopic sample 
in order to avoid overestimation of the \lya\ line flux. 

The rest-frame equivalent widths, $EW_{rest}$, were estimated from 
the line fluxes obtained above and the rest-frame UV continuum flux densities. 
The rest-frame UV continuum flux densities were estimated from the broadband 
photometric data red-ward of the detection IA band. 
The $EW_{rest}$'s before and after the correction for the collected flux
are listed in Table.\ref{tab:spec}. 
Fig.\ref{fig:L-EW} plots $EW_{rest}$ (corrected) against line 
luminosities $L(\lya)$ (corrected).
The errors are estimated by integrating the $1\sigma $ noise level 
of the spectra over the wavelength range where we measured the fluxes.
Six out of the seven objects in our spectroscopic sample have 
large $EW_{rest}$ values exceeding 240\AA. 
The rest-frame \lya\ equivalent width $EW_{rest}= 240\rm\AA$
requires a top-heavy IMF of the slope $\alpha = 0.5$ or 
extremely young age ($\ll 10^7$yrs), 
and thus $EW_{rest}=240\rm\AA$ is thought to be the upper limit for 
normal stellar populations \citep{mr2002}. 
The large equivalent widths of the objects in our spectroscopic sample 
indicate that they are unlikely to be normal starbursts like many other 
LAEs. 
Their $EW_{rest}$ can be still larger, if we take into account the IGM 
absorption of the \lya\ line itself. 
Moreover, because the six objects have continuum 
levels fainter than the $3\sigma$ of the broadband photometric data, 
our estimates of $EW_{rest}$ give a lower limit. 

We also measured the velocity width $\Delta V$ (FWHM) of the \lya\ line. 
The $\Delta V$ value gives important information about their physical 
nature, because $\Delta V$ reflects the kinematical structure of 
the system. 
However, the profile of the \lya\ line could not be significantly resolved 
for any of our objects. 
The measured nominal values are given in Table.\ref{tab:spec}.  
All the objects have $\Delta V$ ranging from 
$\sim 600\kms$ to $\sim 800\kms$, which is comparable to 
the velocity resolutions. 
One object, IB11-59167, shows smaller $\Delta V$ than the 
estimated velocity resolution. 
However, this object is fairly faint, and located in the third 
MOS field where we took integration of only 1.5 hrs; 
the spectrum shows the peak S/N ratio of only $\sim 2$, 
and the $\Delta V$ of this object is not reliable. 

The velocity resolutions estimated from the line widths of the 
nearest sky emission lines are also shown in Table.\ref{tab:spec} 
in the parenthesis. 
These values give rough upper-limits of the intrinsic velocity widths, 
$\Delta V\sim 700-800\kms$. 
We can obtain somewhat stronger upper limits, using 
$\sqrt{(\Delta V + {\rm Err}(\Delta V))^2 - ({\rm resolution})^2}$, 
where ${\rm Err}(\Delta V)$ is the observed error of the velocity width. 
The upper-limits thus estimated were $\simeq 300-800\kms$. 

Although the upper-limits smaller than the velocity resolution are 
not very reliable, 
we can infer from this result that our objects are unlikely 
to be prominent superwind galaxies, which have $\Delta V\gg 1000\kms$ 
(e.g. \citealt{heckman91}). 
Our current resolutions and signal-to-noise ratios (S/N) are not high 
enough to study velocity structures.

\subsection{Luminosity function}
\label{LF}
Our survey shows that 
extended \lya\ sources are quite common in the early universe. 
We divided the number of objects detected (41) by 
our survey volume $\sim 1\times 10^{6}h^{-3}\mbox{Mpc}^3$ (comoving) 
to obtain a number density of $\sim 4\times 10^{-5}h^3\mbox{Mpc}^{-3}$. 
This value is lower than the number density, 
$\sim 4\times 10^{-4}h^3\mbox{Mpc}^{-3}$, obtained from the two 
LABs and the survey volume of S00. 
We derived the luminosity function (LF) of the objects in our sample. 
Fig.\ref{fig:LF} shows the LF of our extended \lya\ sources, 
down to the rest-frame surface brightness limits shown in Table \ref{tab:img}. 
The surface-brightness limits correspond to luminosity limits of 
a few $\times 10^{42}\lum$. 
Since we derived the LF from the raw number of our objects 
(no completeness-correction was made), 
the LF gives a lower limit of the number density. 

The LF in Fig.\ref{fig:LF} shows that the 
number density per unit luminosity is 
$\sim 3-4\times 10^{-5}h^3\,{\rm Mpc^{-3}}/\Delta \log L(\lya)$ 
around $L(\lya)= 1\times 10^{43}\lum$, where $\Delta \log L(\lya)=0.6020$. 
It decreases to $\sim 1\times 10^{-5}h^3\,{\rm Mpc^{-3}}/\Delta \log L(\lya)$ 
around $L(\lya)=3\times 10^{43}\lum$. 
We also plot two data points to show an upper and lower limits of the 
LF (number density) of the two LABs of S00. 
The upper-limit of the number density was derived from the narrowband survey 
volume of S00, which is targeted to the redshift of the two LABs. 
The lower limit was derived from the volume surveyed 
for LBGs at $z\sim 3$ \citep{s03}. 
The survey area of \citet{s03} includes the S00's, but no such objects like 
the two LABs are found within their volume. 
The two brightest LABs have luminosities of $L(\lya)\sim 10^{44}\lum$, 
which is brighter than our objects by an order of magnitude. 
However, their number density is comparable to that of our objects 
at $L(\lya)\simeq 10^{43}\lum$. 

Also plotted in Fig.\ref{fig:LF} are the LFs of 
the original M04 sample and simulated M04 sample. 
The latter was obtained with the imaging simulation 
based on the narrowband images of M04 described in \S\ref{phot}. 
We performed source extraction for the seven simulated images 
with the same manner as our IA images. 
Then we merged the seven samples obtained, 
and calculated the number density by dividing the raw number of 
the objects by $7\times \mbox{(M04's survey volume)}$. 
The number densities derived from the simulated M04 sample are, 
in general,  higher than those of our objects by an order of magnitude. 
The LFs of the original and simulated M04 samples have a tail 
beyond $L(\lya)\simeq 5\times 10^{43}\lum$, 
while no object is found in our sample with such a high luminosity. 
There are no significant differences between the slopes of 
our LF and M04's at $2\times 10^{42}- 2\times 10^{43}\lum$; 
both LFs are flat.

\subsection{Possible physical origins of the extended \lya\ emissions}
\label{origin}
Our extended \lya\ sources have spatially extended strong \lya\ emission 
with: 
(1) typical spatial extents of $\sim 10-15\,\rm kpc$, 
(2) typical \lya\ line luminosities of 
$\sim \mbox{several}\times 10^{42}-1\times 10^{43}$, 
(3) large equivalent widths of $\sim 100-600\rm\AA$, 
and (4) upper limits of the velocity widths of several$\times 10^2\kms$. 
On the other hand, 
(5) they are very faint or compact in continuum emission. 

We can infer possible origins of the extended \lya\ 
emissions from these features. 
Currently proposed scenarios of extended \lya\ emission are; 
({\em a}) cooling radiation of infalling \ion{H}{1} gas, 
({\em b}) resonant scattering of \lya\ photons from the central ionizing 
source by the surrounding \ion{H}{1} gas envelope, or 
({\em c}) shock-heating of the surrounding \ion{H}{1} gas 
by superwind activity. 
We will discuss these scenarios using the features we found. 

\paragraph{({\it a}) Cooling radiation}
\citet{haiman00} and \citet{fardal} predicted that 
primordial \ion{H}{1} gas infalling into a dark matter halo 
would be shock-heated to emit an extended strong \lya\ line 
detectable with current ground-based telescopes. 
In this case, \lya\ photons are originated from 
infalling gas gravitationally heated to $\sim 10^4K$. 
This process does not require UV continuum emission 
to produce \lya\ emission. 
At a temperature of $\sim 10^4K$, 
the \lya\ line is the most effective cooling source, and 
it can release a large part of the gravitational energy. 
The \lya\ line from infalling \ion{H}{1} gas should therefore show 
a much larger equivalent width than that of a usual starburst, 
because there is no significant stellar population. 

According to the simulation of \citet{fardal}, 
the cooling radiation cannot have a spatial extent as large as S00's LABs. 
The predicted spatial extent is consistent with that of our objects, 
$\sim 10-15$ kpc, suggesting that this picture possibly explains 
the extended \lya\ emissions in our sample. 
The \lya\ line luminosities $L(\lya)$ of our objects are also 
consistent with their prediction. 
The system with a \lya\ cooling radiation brighter than $10^{43}\lum$ 
is expected to have high star formation activity, 
and hence $L(\lya)$ becomes larger than $10^{44}\lum$. 
The luminosity range of our objects is consistent with 
that of cooling-radiation-dominated systems. 
Their simulation also predicts that the velocity width 
cannot significantly exceed the circular velocity of the system, 
typically $\sim 300\kms$ in half-width-half-maximum ($\sim 600\kms$ in FWHM), 
even if the effect of resonant scattering is taken into account. 
Our results of $\Delta V \lesssim 700-800\kms$ (FWHM) therefore 
do not conflict with this picture. 
Since both the $L(\lya)$ and $\Delta V$ are scaled to the mass of the 
system, the correlation between $L(\lya)$ and $\Delta V$ should 
be a diagnostic of this scenario. However, our current spectral data 
do not have a sufficient quality, and this measurement should be done with 
future spectroscopy with a higher resolution and S/N ratio. 

\paragraph{({\it b}) Resonant scattering of \lya\ photons}
The \ion{H}{1} gas envelope surrounding an ionizing source can 
make the \lya\ emission spatially extended by resonant scattering 
\citep{haiman00}. 
\lya\ photons from the ionizing source is absorbed by surrounding 
\ion{H}{1} gas, and re-emitted toward different directions. 
This process leads to a spatially extended \lya\ emission. 
Since the spatial extents of our sample are 
only $\sim 10-15\,\rm kpc$, this reprocess by the \ion{H}{1} 
gas surrounding the central starburst/AGN is a possible 
explanation. 

If the ionizing source is an AGN, the observed large equivalent widths 
of the \lya\ line can be easily explained. 
There is actually an extended \lya\ source which is best explained 
by resonant scattering \citep{weidinger04, weidinger05}. 
For known \lya\ nebulae associated with AGNs, however, 
the velocity widths are generally greater than $\sim 1000\kms$ 
\citep{vanojik}. 
The spatial extents of such nebulae are also larger than the objects  
in our sample in general ($\gtrsim 100\,\rm kpc$: \citealt{heckman91a}). 
Therefore, if the resonant scattering of AGN-originated \lya\ photons 
is the case, our objects should be weak AGNs. 
Powerful AGNs are unlikely. 

If the ionizing source is a starburst, on the other hand, 
their large equivalent widths require extreme cases. 
\citet{mr2002} showed that extremely young age ($\ll 10^7\rm yrs$) 
or top-heavy IMFs like $\alpha = 0.5$ can yield $EW_{rest}$ values larger 
than 300\AA. 
This suggests that the objects in our sample are unlikely 
to be normal starbursts. 
An extremely young stellar population may explain 
the observed large equivalent widths of our objects. 
In fact, most of the high-$z$ LAEs can be understood as actively 
star-forming galaxies even if $EW_{rest}$ is larger than 240\AA\ 
(e.g. \citealt{wang04}). 
Another extreme case is that the continuum emission from 
the central starburst is obscured by dust. 
In this case, both the continuum and \lya\ line should suffer from 
the absorption. 
However, if the surrounding gas envelope is spatially more extended than 
the compact starburst region, then 
the line emitter (gas) suffers from less absorption than the 
ionization source (starburst). 
Therefore, photo-ionization by starbursts is a possible explanation. 

\paragraph{({\it c}) Superwind}
For starburst-induced superwind galaxies known to date, 

The measured velocity widths $\Delta V$ of starburst-induced superwind 
galaxies known to date range from 
several $100\kms$ to $\gtrsim 10^{3}\kms$. 
Because upper-limits for our objects are $\Delta V \sim 700-800\kms$, 
our objects are unlikely to be galaxies like known powerful superwinds. 
The spatial extents of our objects are also smaller than 
the typical powerful superwind source, Arp220, 
which has ionized gas extended up to several tens of kpc \citep{tts98}. 

However, we cannot rule out this scenario from our current data. 
The upper-limit of $\Delta V\lesssim 800\kms$ for our sample could be 
consistent with superwinds with relatively small $\Delta V$. 
There are two possible diagnostics of the superwind scenario. 
One is the velocity structure of the \lya\ emission. 
If the \lya\ nebula is shock-heated by the outflow activity of the 
galaxy, the \lya\ line must show characteristic velocity structure 
of an expanding bubble. 
Another is the existence of \ion{C}{4} line emission. 
If the starburst-driven superwind is the origin of the 
extended \lya\ emission, the primordial gas surrounding the system 
would be in some part chemically enriched. 
This enrichment would be detected as an extended \ion{C}{4} emission 
with a strength of typically 7--10\% of the \lya\ line \citep{heckman91}. 
However, both diagnostics require spectra of a higher resolution and 
S/N ratio than our current data. 

\section{Conclusions}
\label{conclusion}
We have carried out a systematic survey for spatially extended \lya\ 
sources which are faint and/or compact in rest-frame UV continuum, 
using seven intermediate-band images taken with Subaru/Suprime-Cam. 
We identified 41 objects in the redshift range of 
$3.24\lesssim z\lesssim 4.95$. 
The lower limit of the number density was calculated to be 
$\sim 4\times 10^{-5}h^3\rm \rm Mpc^{-3}$ over a surveyed volume of 
$\sim 1\times 10^{6}h^{-3}\rm Mpc^3$, down to $L(\lya)\sim 10^{42}\lum$. 
These results show that this kind of extended \lya\ sources 
are quite common in the early universe. 
A comparison of the luminosity function (LF) between our sample and M04's 
showed that 
the number density of our objects is smaller than those of M04's sample 
by an order of magnitude. 
No objects are found in our sample with luminosities as bright as 
$\gtrsim \mbox{several}\times 10^{43}\lum$. 
This result shows that the largest and brightest objects 
like the two LABs of S00 are found only in highly overdense regions.

The \lya\ line emission components of the objects in our sample have 
typical sizes of $\sim 10-15\,\rm kpc$, and luminosities of 
several $\times 10^{42}\lum$. 
Follow-up spectroscopy suggested that our sample suffers from little 
contamination. The objects in our sample have fairly large equivalent widths 
of \lya\ line ($EW_{rest}\sim 100 - 600\rm\AA$). 
Six out of the seven objects with spectroscopy have $EW_{rest}\ge 240\rm\AA$; 
they cannot be normal starbursts. 
The ionizing source of the extended \lya\ emission of our objects  
is likely to be either a non-stellar process of a forming galaxy, 
or a stellar UV radiation of an extremely young ($\ll 10^{7}$ yrs) or 
obscured starburst. 

\acknowledgements
We thank Dr. Toru Yamada and Dr. Masanori Iye for their giving 
helpful advice and continuous support for preparing the manuscript. 
We thank Dr. Yuichi Matsuda for providing the narrowband imaging data, 
and giving us an important suggestion. 
We thank SXDS team for a fruitful discussion. 
We thank the anonymous referee for the valuable 
comments which have greatly improved the paper.


\clearpage


\begin{deluxetable}{cccccc}
\tablecaption{Summary of the imaging data\label{tab:img}}
\tabletypesize{\scriptsize}
\tablewidth{0pt}
\tablehead{
\colhead{Band}&
\colhead{Exposure}&
\colhead{$\rm m_{lim}(3\sigma)$\tablenotemark{a}}&
\colhead{$\Sigma_{lim}(2\sigma)$\tablenotemark{b}}&
\colhead{$N(>3\sigma)$\tablenotemark{c}}&
\colhead{Redshift}\\
\colhead{} &
\colhead{[sec]} &
\colhead{[mag]}&
\colhead{[$\rm ergs\, cm^{-2}s^{-1}arcsec^{-2}$]}&
\colhead{} &
\colhead{of \lya}
}
\startdata
IA527 &  5280 & 26.8 & $1.27\times 10^{-15}$ & 67405 & $3.34\pm 0.10$\\
IA574 &  7200 & 26.5 & $2.37\times 10^{-15}$ & 57824 & $3.72\pm 0.11$\\
IA598 &  6720 & 26.5 & $2.80\times 10^{-15}$ & 60219 & $3.93\pm 0.12$\\
IA624 & 10560 & 26.7 & $2.56\times 10^{-15}$ & 69571 & $4.12\pm 0.12$\\
IA651 &  6500 & 26.7 & $2.76\times 10^{-15}$ & 71365 & $4.35\pm 0.13$\\
IA679 & 10560 & 26.8 & $3.11\times 10^{-15}$ & 70228 & $4.58\pm 0.14$\\
IA709 & 11520 & 26.6 & $3.86\times 10^{-15}$ & 68535 & $4.82\pm 0.13$\\
$B$&  18000& 28.2 & --- & 115354 & ---\\
$R$&  12000& 27.4 & --- & 95230 & ---\\
$i'$& 13200& 27.0 & --- & 85198 & ---\\
$z'$& 5700&  25.8 & --- & 51599 & ---
\enddata
\tablenotetext{a}{The 3-$\sigma$ limiting magnitude measured with 
a $2''\phi$ circular aperture. }
\tablenotetext{b}{The threshold of the source extraction in rest-frame surface brightness of the \lya\ line.}
\tablenotetext{c}{The number of objects brighter than $3\sigma$.}
\end{deluxetable}


\begin{deluxetable}{cccc}
\tablecaption{Color selection criteria of \lya\ sources\label{tab:select}}
\tabletypesize{\small}
\tablewidth{0pt}
\tablehead{
\multicolumn{2}{c}{Detected band}&
\colhead{\lya-excess}&
\colhead{$B$-nondetection}\\
\colhead{(\lya)} &
\colhead{(Cont)} &
\colhead{}&
\colhead{or dropout}
}
\startdata
IA527& $R$& $R-\mbox{MAG\_AUTO(IA527)}>0.75$ & $B>28.7$ or $B-R>0.6$\\
IA574& $R$& $R-\mbox{MAG\_AUTO(IA574)}>0.75$ & $B>28.7$ or $B-R>1.4$\\
IA598& $i'$& $i'-\mbox{MAG\_AUTO(IA598)}>0.75$ & $B>28.7$ \\
IA624& $i'$& $i'-\mbox{MAG\_AUTO(IA624)}>0.75$ & $B>28.7$ \\
IA651& $i'$& $i'-\mbox{MAG\_AUTO(IA651)}>0.75$ & $B>28.7$ \\
IA679& $i'$& $i'-\mbox{MAG\_AUTO(IA679)}>0.75$ & $B>28.7$ \\
IA709& $z'$& $z'-\mbox{MAG\_AUTO(IA709)}>0.75$ & $B>28.7$ 
\enddata
\tablecomments{
The second column (Cont) means the band where we measure the continuum 
levels, i.e. the nearest red-ward broadband of the \lya\ line. 
Magnitudes are measured with a $2''\phi$ aperture unless otherwise noted. 
The criteria for spatial extent are common for all the IA bands; 
FWHM(IA)$>1''.4$ and FWHM(Cont)$<1''.2$. 
}
\end{deluxetable}


\begin{deluxetable}{lcccccccccccccccc}
\tablecaption{Photometric properties of extended \lya\ sources\label{tab:phot}}
\tabletypesize{\scriptsize}
\rotate
\tablewidth{0pt}
\setlength{\tabcolsep}{1mm}
\tablehead{
\colhead{}&
\colhead{$\alpha$}&\colhead{$\delta$}&
\colhead{}&\colhead{}&\colhead{}&\colhead{}&\colhead{}&
\colhead{}&\colhead{}&\colhead{}&\colhead{}&\colhead{}&\colhead{}&
\colhead{auto}&
\multicolumn{2}{c}{FWHM [$''$]}\\
\colhead{Object ID} &
\multicolumn{2}{c}{(J2000)}&
\colhead{$B$} & \colhead{$R$} & \colhead{$i'$} & \colhead{$z'$} &
\colhead{IA527} & \colhead{IA574} & \colhead{IA598} & \colhead{IA624} &
\colhead{IA651} & \colhead{IA679} & \colhead{IA709} &
\colhead{(IA)} &
\colhead{(IA)} & \colhead{(Cont)}
}
\startdata
 IB08-2031&  2:17:56.66& -5:37:47.7& 
28.21& {\bf 27.45}& 27.41& 27.00& {\bf 25.83}&
27.70& 27.56& 27.90& 26.71& 27.64& 27.80& 25.24&  $1.71$& $1.46$\\
IB08-20047&  2:18:43.50& -5:33:48.0& 
27.06& {\bf 26.36}& 26.33& 26.11& {\bf 25.93}&
27.17& 26.81& 26.18& 26.17& 26.59& 26.49& 25.21&  2.15& 1.14\\
IB08-27691&  2:18:59.40& -5:31:58.8& 
29.40& {\bf 28.15}& 27.47& 27.00& {\bf 25.77}&
27.70& 27.52& 27.90& 27.82& 27.24& 27.53& 25.65&  1.42& 0.93\\
IB08-28421&  2:18:20.07& -5:31:49.9& 
27.32& {\bf 26.45}& 26.50& 27.00& {\bf 25.94}&
26.88& 26.89& 26.89& 26.81& 26.82& 26.34& 25.62&  1.48& 1.09\\
IB08-36255&  2:17:20.80& -5:29:58.6& 
26.59& {\bf 25.85}& 25.87& 25.69& {\bf 25.58}&
26.16& 26.11& 26.03& 25.90& 25.89& 25.70& 24.73&  1.55& 1.16\\
IB08-86220&  2:18:28.32& -5:18:11.9& 
27.20& {\bf 26.51}& 26.48& 27.00& {\bf 25.76}&
26.40& 26.72& 26.43& 26.68& 27.20& 27.10& 25.40&  1.63& 1.20\\
IB10-17108&  2:16:58.03& -5:34:19.1& 27.16& {\bf 25.65}& 25.65& 25.72& 26.63&
{\bf 25.07}& 25.49& 25.75& 25.63& 25.84& 25.74& 24.83&  1.41& 1.09\\
IB10-17580&  2:18:29.03& -5:34:08.8& 28.19& {\bf 26.39}& 26.28& 25.95& 28.00&
{\bf 25.76}& 27.09& 27.13& 26.62& 27.10& 26.65& 25.28&  2.15& 1.16\\
IB10-32162&  2:16:56.88& -5:30:29.5& 27.53& {\bf 25.98}& 26.30& 26.14& 26.85&
{\bf 25.22}& 25.73& 26.00& 26.12& 26.14& 26.17& 24.90&  1.68& 1.14\\
IB10-54185&  2:17:59.46& -5:25:07.2& 28.16& {\bf 26.44}& 26.42& 26.61& 27.31&
{\bf 25.61}& 25.98& 26.55& 26.42& 26.59& 26.26& 25.13&  1.46& 1.08\\
IB10-79754&  2:19:02.00& -5:18:54.7& 29.40& {\bf 28.60}& 28.20& 26.46& 28.00&
{\bf 25.94}& 27.70& 27.90& 27.90& 28.00& 27.36& 25.87&  1.48& 2.09\\
IB10-90651&  2:17:43.32& -5:16:12.0& 29.40& {\bf 27.75}& 27.79& 27.00& 26.96&
{\bf 25.84}& 27.23& 27.90& 27.90& 27.39& 26.66& 25.50&  2.42& 2.87\\
 IB11-3727&  2:19:06.14& -5:37:29.1& 29.40& 28.60& {\bf 28.20}& 27.00& 28.00&
27.51& {\bf 25.94}& 27.90& 27.55& 28.00& 27.80& 26.04&  2.47& 0.00\\
 IB11-5427&  2:18:30.81& -5:37:10.7& 29.37& 26.16& {\bf 26.61}& 27.00& 26.66&
27.70& {\bf 25.86}& 25.76& 26.16& 26.93& 26.29& 25.62&  1.49& 1.18\\
 IB11-6145&  2:16:53.46& -5:36:58.8& 29.40& 26.36& {\bf 27.17}& 26.65& 27.12&
27.70& {\bf 25.63}& 26.21& 27.46& 26.99& 27.80& 25.28&  2.27& 1.38\\
IB11-59167&  2:17:10.19& -5:23:47.2& 29.40& 26.71& {\bf 27.59}& 27.00& 27.43&
26.92& {\bf 25.80}& 27.90& 27.77& 27.63& 27.37& 25.54&  2.08& 2.29\\
IB11-66309&  2:18:36.55& -5:21:52.3& 29.40& 26.62& {\bf 26.92}& 27.00& 27.32&
27.68& {\bf 25.77}& 26.54& 26.95& 27.02& 27.80& 25.60&  1.92& 1.10\\
IB11-80344&  2:17:44.67& -5:18:15.0& 29.08& 26.70& {\bf 27.03}& 26.95& 28.00&
27.59& {\bf 25.53}& 26.85& 27.29& 26.75& 26.96& 25.24&  2.07& 1.88\\
IB11-89537&  2:17:45.31& -5:15:53.2& 29.40& 26.65& {\bf 27.25}& 27.00& 28.00&
27.70& {\bf 25.80}& 25.58& 27.90& 27.49& 27.70& 25.52&  1.49& 1.23\\
IB11-101786&  2:17:47.30& -5:13:05.2& 29.40& 27.22& {\bf 27.01}& 26.39& 28.00&
27.70& {\bf 25.97}& 27.90& 27.68& 28.00& 27.80& 25.92&  1.73& 2.12\\
 IB12-1092&  2:19:06.27& -5:37:53.3& 29.40& 26.71& {\bf 26.52}& 27.00& 27.00&
27.70& 26.28& {\bf 25.81}& 26.16& 26.90& 26.60& 25.66&  1.86& 1.14\\
IB12-21989&  2:17:13.33& -5:32:56.5& 29.40& 25.12& {\bf 25.88}& 25.72& 28.00&
27.48& 26.66& {\bf 25.71}& 26.47& 26.15& 26.02& 25.06&  1.48& 1.13\\
IB12-30834&  2:17:55.95& -5:30:53.2& 29.40& 27.10& {\bf 27.09}& 27.00& 28.00&
27.70& 27.70& {\bf 25.82}& 27.84& 27.76& 27.80& 25.71&  1.47& 1.11\\
IB12-48320&  2:16:55.87& -5:26:36.9& 29.07& 26.59& {\bf 27.35}& 25.82& 28.00&
27.70& 26.05& {\bf 25.82}& 27.90& 26.82& 27.56& 25.35&  1.61& 2.53\\
IB12-58572&  2:17:49.08& -5:24:11.4& 29.40& 27.16& {\bf 26.82}& 27.00& 26.87&
27.70& 26.52& {\bf 25.97}& 26.84& 26.64& 27.34& 25.32&  1.57& 1.17\\
IB12-71781&  2:17:38.93& -5:20:57.8& 29.16& 26.68& {\bf 26.77}& 26.71& 27.79&
27.70& 26.89& {\bf 25.96}& 27.32& 27.34& 26.49& 25.67&  1.55& 1.15\\
IB12-81981&  2:18:14.69& -5:18:32.2& 29.40& 26.85& {\bf 27.84}& 27.00& 26.88&
27.70& 27.70& {\bf 25.65}& 26.87& 26.79& 27.62& 25.27&  1.77& 2.80\\
  IB13-999&  2:18:56.49& -5:37:54.6& 29.24& 27.42& {\bf 28.20}& 27.00& 27.76&
27.70& 27.70& 27.90& {\bf 25.98}& 28.00& 27.80& 25.87&  2.00& 2.34\\
IB13-20748&  2:18:41.91& -5:33:19.4& 29.40& 26.57& {\bf 27.21}& 26.30& 28.00&
27.54& 27.70& 27.90& {\bf 25.60}& 26.47& 27.30& 25.20&  1.58& 2.00\\
IB13-28369&  2:18:39.14& -5:31:26.7& 29.40& 26.70& {\bf 27.49}& 27.00& 28.00&
27.70& 27.52& 27.34& {\bf 25.59}& 27.61& 27.80& 24.98&  1.61& 2.48\\
IB13-62009&  2:17:28.52& -5:23:17.5& 29.40& 26.83& {\bf 27.08}& 27.00& 27.16&
27.32& 27.36& 26.24& {\bf 25.95}& 27.77& 27.44& 25.53&  2.23& 2.85\\
IB13-95072&  2:17:00.33& -5:15:19.2& 29.40& 26.77& {\bf 27.26}& 27.00& 28.00&
27.70& 27.70& 27.90& {\bf 25.49}& 28.00& 27.80& 25.31&  1.44& 1.86\\
IB13-96047&  2:18:13.30& -5:15:05.0& 29.40& 26.16& {\bf 25.97}& 25.85& 28.00&
26.99& 27.65& 27.90& {\bf 25.80}& 25.85& 26.04& 25.16&  1.54& 1.18\\
IB13-104299&  2:18:21.06& -5:13:24.8& 29.17& 26.56& {\bf 27.45}& 26.71& 28.00&
26.62& 27.70& 27.90& {\bf 25.87}& 26.78& 26.19& 25.48&  3.15& 3.15\\
IB14-47257&  2:18:01.94& -5:25:25.0& 29.40& 26.62& {\bf 26.29}& 26.65& 28.00&
27.56& 27.11& 27.54& 27.54& {\bf 25.93}& 26.44& 25.14&  1.44& 1.19\\
IB14-52102&  2:18:00.10& -5:24:10.3& 29.40& 26.21& {\bf 26.33}& 26.06& 27.54&
26.67& 26.78& 27.03& 26.01& {\bf 25.39}& 25.83& 25.15&  1.43& 1.11\\
IB14-62116&  2:17:58.21& -5:21:35.4& 28.82& 26.08& {\bf 25.88}& 26.09& 28.00&
27.70& 26.57& 27.69& 27.90& {\bf 25.14}& 26.09& 24.74&  1.50& 1.14\\
IB15-66451&  2:18:33.92& -5:22:09.3& 29.40& 26.47& 25.70& {\bf 25.87}& 28.00&
27.70& 27.70& 27.35& 27.71& 28.00& {\bf 25.23}& 24.79&  1.58& 1.58\\
IB15-94328&  2:16:55.31& -5:15:22.0& 29.17& 26.85& 25.73& {\bf 25.95}& 28.00&
27.70& 27.70& 27.24& 27.22& 27.23& {\bf 25.27}& 25.09&  1.52& 2.40
\enddata
\tablecomments{
Magnitudes were measured with a $2''\phi$ circular aperture, 
except the column ``auto''.  
For objects fainter than the $1\sigma$ flux limit, 
the $1\sigma$ limiting magnitudes are shown instead of 
the SExtractor's outputs. 
The ``auto'' columon shows MAG\_AUTO measured in the detected IA band 
(i.e. \lya\ line). 
The FWHM means the spatial extent in FWHM measured on the image (in arcsec). 
The (IA) is for the detected IA band, 
and (Cont) is for the red-ward continuum band. 
The magnitudes in the red-ward continuum band and the detected IA band 
are written in bold characters. 
The $\alpha$ and $\delta$ are the right ascension 
and declination (J2000) based on the beta-version astrometry 
given in the SXDS broadband data. 
}
\end{deluxetable}


\begin{deluxetable}{cccccccc}
\tablecaption{Spectral properties of extended \lya\ sources\label{tab:spec}}
\tabletypesize{\scriptsize}
\tablewidth{0pt}
\tablehead{
\colhead{Object ID} &
\colhead{$\lambda_c$\tablenotemark{a}} &
\colhead{$z$} &
\colhead{$F(\lya)$\tablenotemark{b}} &
\colhead{$L(\lya)$\tablenotemark{b}} &
\colhead{$EW_{rest}$\tablenotemark{b}} &
\colhead{$\Delta V$ (resolution)\tablenotemark{c}} &
\colhead{$\Delta V_{upper}$\tablenotemark{d}}\\
\colhead{\ } &
\colhead{[\AA]} &
\colhead{} &
\colhead{[\fluxunit]} &
\colhead{[\lum]} &
\colhead{[\AA]} &
\colhead{[\kms]} &
\colhead{[\kms]}
}
\startdata
IB08-86220&  5235&  3.31&  
$1.2\pm 0.6\; (3.0)\times 10^{-17}$&  $1.2(3.0)\times 10^{42}$&
    44 (110)& 748 (750) & 550\\
IB10-90651&  5692&  3.68&  
$1.6\pm 0.5\; (4.0)\times 10^{-17}$&  $2.0(5.0)\times 10^{42}$&
   164 (410)& 795 (740) & 440\\
IB11-59167&  6005&  3.94&  
$8.6\pm 7\; (22)\times 10^{-18}$&  $1.3(3.2)\times 10^{42}$&
    75 (188)& 627 (700) & 300\\
IB11-80344&  5943&  3.89&  
$2.6\pm 0.6\; (6.5)\times 10^{-17}$&  $3.8(9.5)\times 10^{42}$&
   156 (390)& 795 (780) & 770\\
IB11-89537&  6110&  4.02&  
$2.1\pm 0.4\; (5.2)\times 10^{-17}$&  $3.2(8.0)\times 10^{42}$&
   136 (340)& 779 (780) & 550\\
IB12-48320&  6123&  4.04&  
$1.6\pm 0.5\; (4.0)\times 10^{-17}$&  $2.6(6.5)\times 10^{42}$&
   132 (330)& 695 (700) & 450\\
IB12-81981&  6225&  4.12&  
$1.6\pm 0.5\; (4.0)\times 10^{-17}$&  $2.7(6.8)\times 10^{42}$&
   179 (448)& 789 (670)& 680
\enddata
\tablenotetext{a}{wavelength of the line center.}
\tablenotetext{b}{
Indicated in the parenthesis is the value corrected for the 
flux loss with the $0''.8$-wide slit.}
\tablenotetext{c}{Velocity resolutions are shown in the parenthesises.}
\tablenotetext{d}{Upper limits of the velocity widths.}
\end{deluxetable}


\begin{figure}
\begin{center}
\plotone{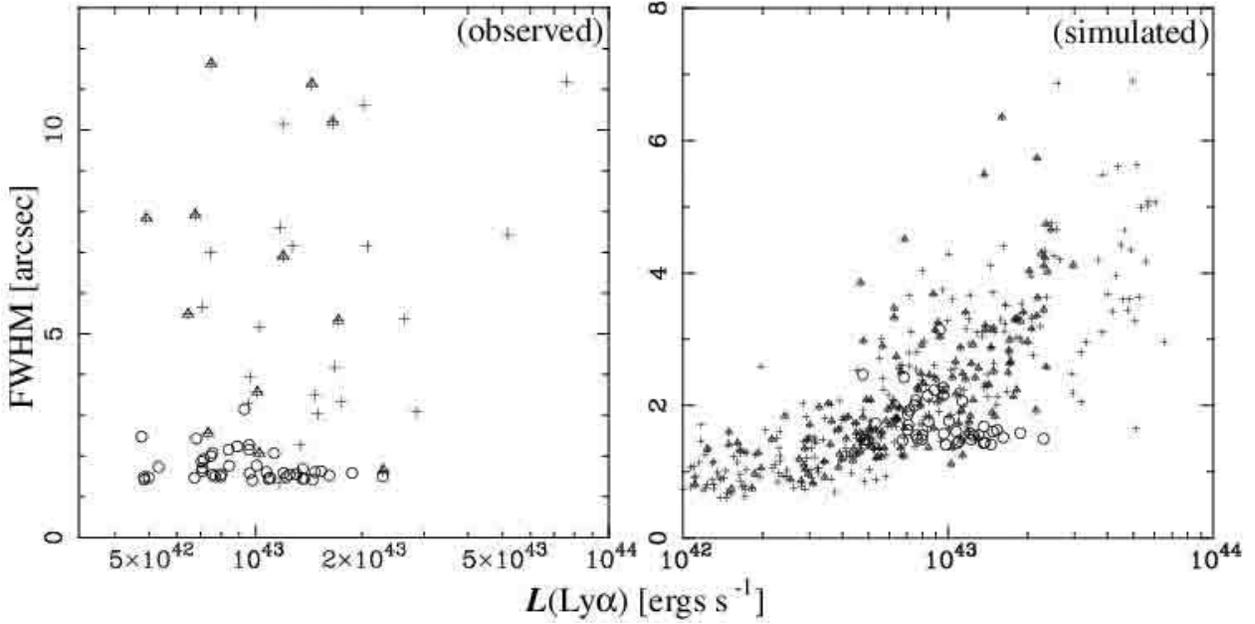}
\caption{Physical sizes in FWHM plotted against \lya\ line luminosity. 
For both panels, the open circles represent the 41 candidates in our sample. 
For M04's sample, the crosses represent the all objects, 
and the triangles represent 12 objects which have large 
equivalent widths ($EW_{rest}\ge 54\rm\AA$). 
The left panel shows a comparison with the M04 sample, using the original 
values given in M04's catalogue. 
The right panel shows a comparison with the simulated M04 sample 
obtained with the IA imaging simulation described in \S\ref{phot}. 
}
\label{fig:l-fwhm}
\end{center}
\end{figure}

\begin{figure}
\begin{center}
\plotone{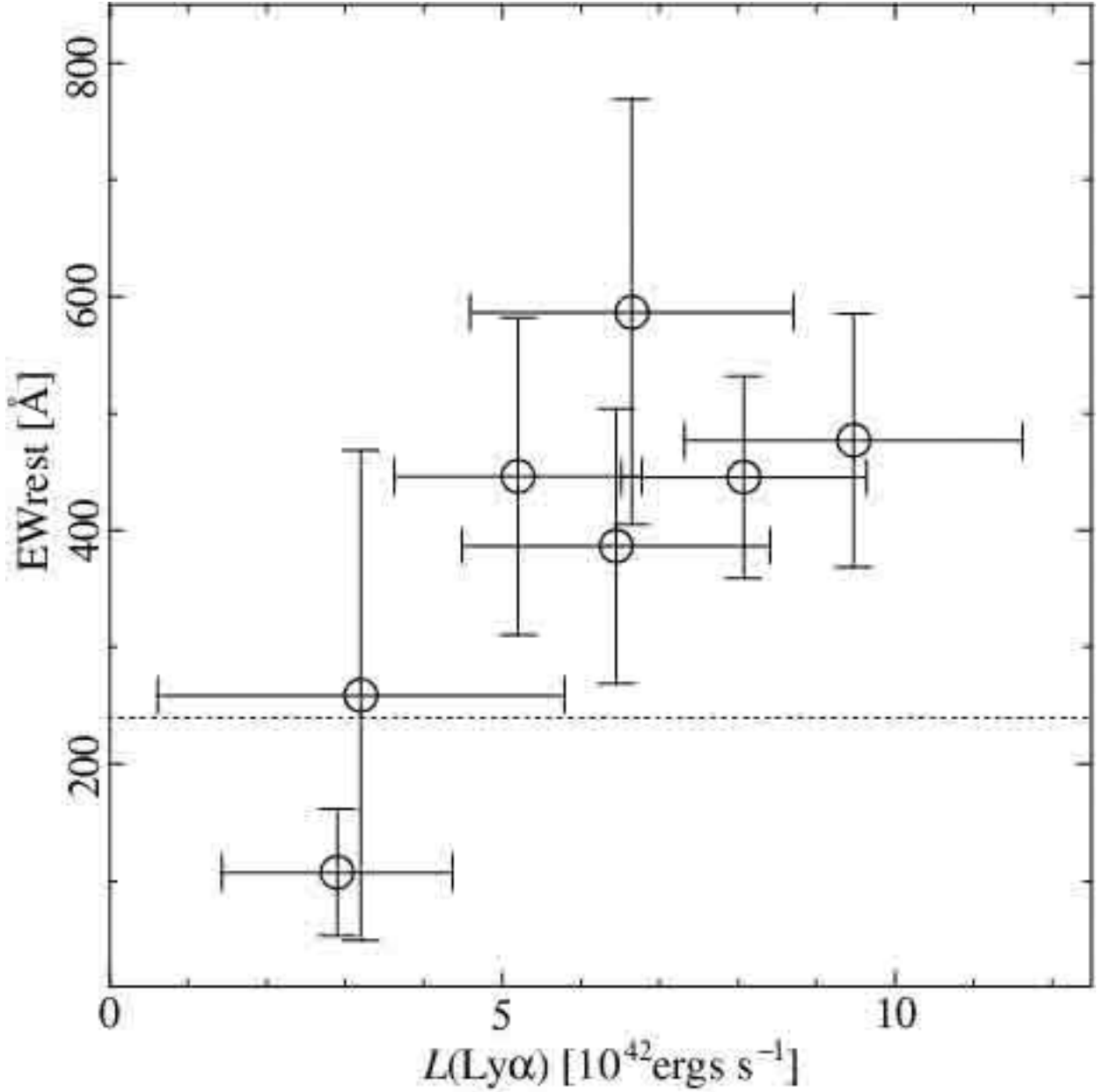}
\caption{Rest-frame equivalent width of \lya\ line 
as a function of \lya\ luminosity for our spectroscopic sample. 
The horizontal dotted line shows 240\AA, which is an 
upper-limit for \lya\ emission of stellar origin \citep{mr2002}.
The error bars were derived from the rms noise level of the spectral data. 
}
\label{fig:L-EW}
\end{center}
\end{figure}

\begin{figure}
\begin{center}
\plotone{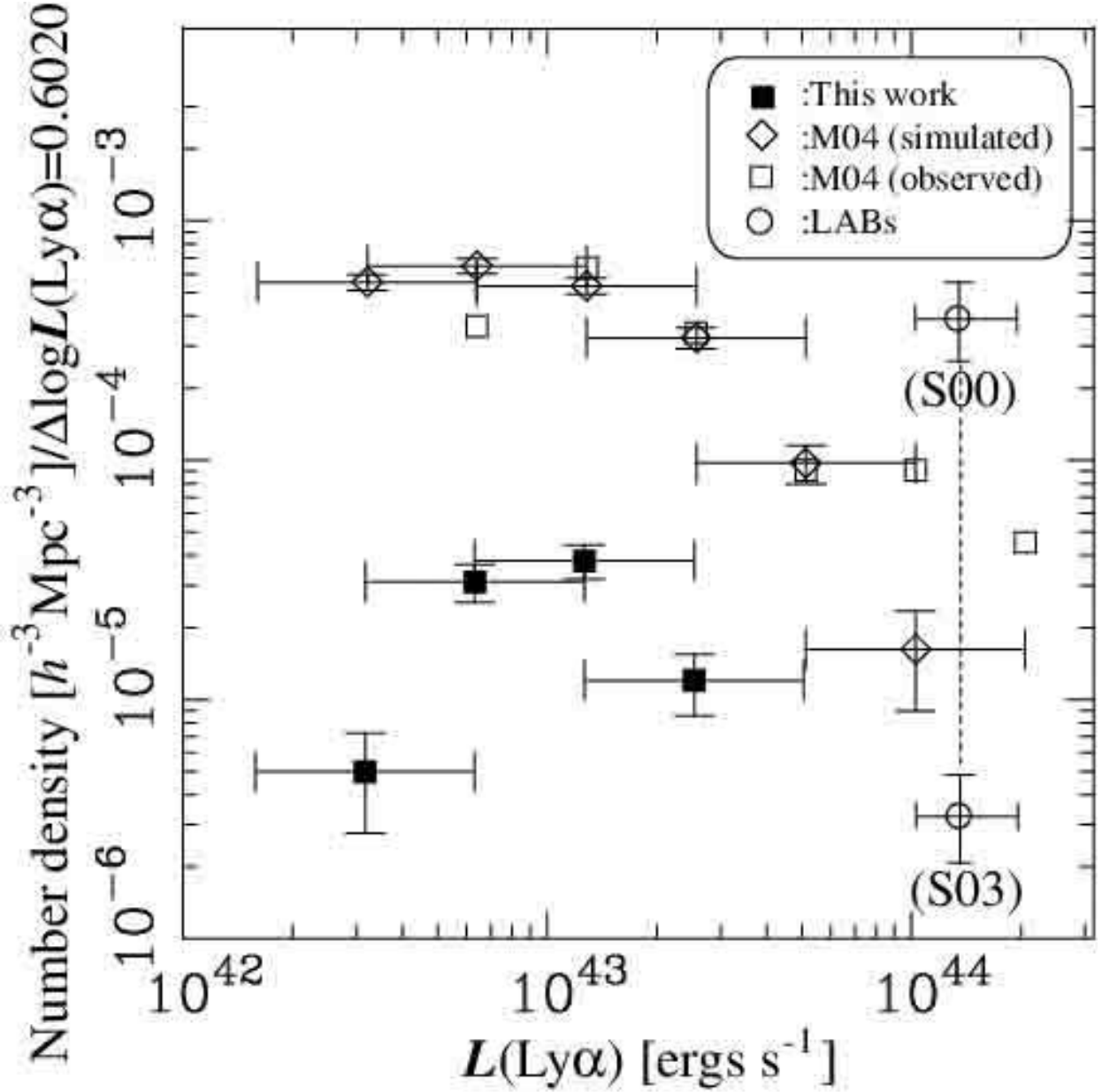}
\caption{Luminosity functions (LFs) of extended \lya\ sources. 
The filled squares show the LF of our sample. 
The open squares indicate the LF of M04 the sample (original values). 
The open diamonds are the LF of the simulated M04 sample, 
which was obtained with the imaging simulation 
described in \S\ref{phot} (see the text). 
The two open circles connected by the dashed line at the right 
end correspond to the number density of the two LABs, 
i.e., two divided by the survey volume of S00 (marked ``(S00)'') and 
that of \citet{s03} (marked ``(S03)''). 
For all the points, 
the horizontal error bars represent the size of the luminosity bin, 
and the vertical ones represent the Poisson error. 
No completeness correction was made for any of the points. 
}
\label{fig:LF}
\end{center}
\end{figure}


\end{document}